\documentclass[%
 prx,
rsi,%
 amsmath,amssymb,
 reprint,%
superscriptaddress
]{revtex4-1}
\usepackage{graphicx}
\usepackage{dcolumn}
\usepackage{bm}
\usepackage{color}
\usepackage{amssymb}
\usepackage{amsmath}
\usepackage{mathtools}
\usepackage{tabularx}
\usepackage{booktabs}
\usepackage{tikz}
\usepackage{placeins}

\begin{document}
\title{Active spin model for cell assemblies on 1D substrates}

\author{Harshal Potdar}
\affiliation{Department of Physics, Savitribai Phule Pune University, Pune, 411007, India.}
\author{Ignacio Pagonabarraga}
\affiliation{Departament de F\'{\i}sica de la Mat\`eria Condensada, Universitat de Barcelona, C. Mart\'{\i} i Franqu\`es 1, E08028 Barcelona, Spain.}
\affiliation{UBICS University of Barcelona Institute of Complex Systems,  Mart\'{\i} i Franqu\`es 1, E08028 Barcelona, Spain.}
\author{Sudipto Muhuri}
\email{sudiptomuhuri@uohyd.ac.in}
\affiliation{Department of Physics, Savitribai Phule Pune University, Pune, 411007, India.}
\affiliation{School of Physics, University of Hyderabad, Hyderabad, 500046, India.}
\date{\today}
            
\begin{abstract}
The experimental use of micropatterned quasi-1D substrates has emerged as an useful experimental tool to study the nature of cell-cell interactions and gain insight on collective behaviour of cell colonies. Inspired by these experiments, we propose an active spin model to investigate the emergent properties of the cell assemblies. The lattice gas model incorporates the interplay of self-propulsion, polarity directional switching, intra-cellular attraction, and contact Inhibition Locomotion (CIL). In the absence of vacancies, which corresponds to a confluent cell packing on the substrate, the model reduces to an equilibrium spin model which can be solved exactly. In the presence of vacancies, the clustering is controlled by a dimensionless Pecl\'et Number, $Q$ - the ratio of magnitude of self-propulsion rate and directional switching rate of particles. In the absence of CIL interactions, we invoke a mapping to Katz-Lebowitz-Spohn model to determine an exact analytical form of the cluster size distribution in the limit  $Q \ll 1$. In the limit of $Q \gg 1$, the cluster size distribution exhibits an universal scaling behaviour (in an approximate sense), such that the distribution function can be expressed as a scaled function of Q, particle density and CIL interaction strength. 
We characterize the phase behaviour of the system in terms of contour plots of average cluster size.
 The average cluster size exhibit a non-monotonic dependence on CIL interaction strength, attractive interaction strength, and self-propulsion.   
\end{abstract}
\maketitle

\section{Introduction}

Collective cell behavior is the main driver for a wide class of intracellular processes ranging from embryonic development, cancer progression, to cell migration in tissues \cite{igna-pnas,trepat,ignaref2,cancer,FEBS}. Unraveling the underlying physical mechanisms at play which lead to the emergence of collective behaviour of cells and their collective migration is crucial to understanding the nature of cellular organization in tissues and its potential implications for a wide variety of relevant processes such as wound healing, tissue morphogenesis, and cancer progression \cite{trepat,cancer,FEBS}. 

A characteristic feature of all cells is their ability to self-propel \cite{trepat,igna-pnas,ignaref2,levineref14}. Viewed from the perspective of active matter physics, collective cellular organization is a paradigmatic example of emergent collective behaviour arising out of the complex interaction of self-propelled {\it active} entities at microscopic level \cite{sriram-rev,igna-pnas}. The nature of interactions between individual cells can be grouped into two types: a) Interactions which depend on the position coordinates of the cells and b) Interactions which depend on the orientation or polarity of the cells. For many cell types, such as epithelial cells, neighboring cells tend to form adhesions \cite{cadherin,trepat,igna-pnas}. Typically, cell-cell adhesion is mediated by transmembrane protein complexes which link the actomyosin cortices of the neighbouring cells \cite{cadherin,trepat}. Additionally, neighbouring cells also experience a short repulsion on account of cytoskeletal elasticity of the individual cells \cite{trepat}. Both these interactions result in an effective short range attraction, and a short range repulsion which depends on the position coordinates of the neighbouring cells. In addition to these effects for large class of cells, it has been observed that there is an inherent propensity for the cells to align their individual polarity (self propulsion direction) away from each other when they come in contact \cite{cil1, cil2, cil-plos}. This phenomenon is also referred to as contact inhibition locomotion (CIL) \cite{cil1, cil2, cil-plos, levine-pnas}. The CIL interaction arises because when two cells collide, the cell front adheres to the colliding cell, obstructing further cell movement \cite{cil1,cil2,cil-plos}. This triggers a process of repolarization of the cell cytoskeleton that eventually results in cells separating and moving away from each other \cite{cil1,cil2,cadherin}. 
In addition, cells exhibit a propensity to align their polarity with each other upon contact \cite{trepat}. Together, all these interactions manifest themselves in the collective behaviour of cell assemblies and are instrumental in shaping the functional and organizational characteristics of cell assemblies \cite{igna-pnas}.  

\begin{figure*}
	\centering
	\includegraphics[width = \linewidth, height = 10cm]{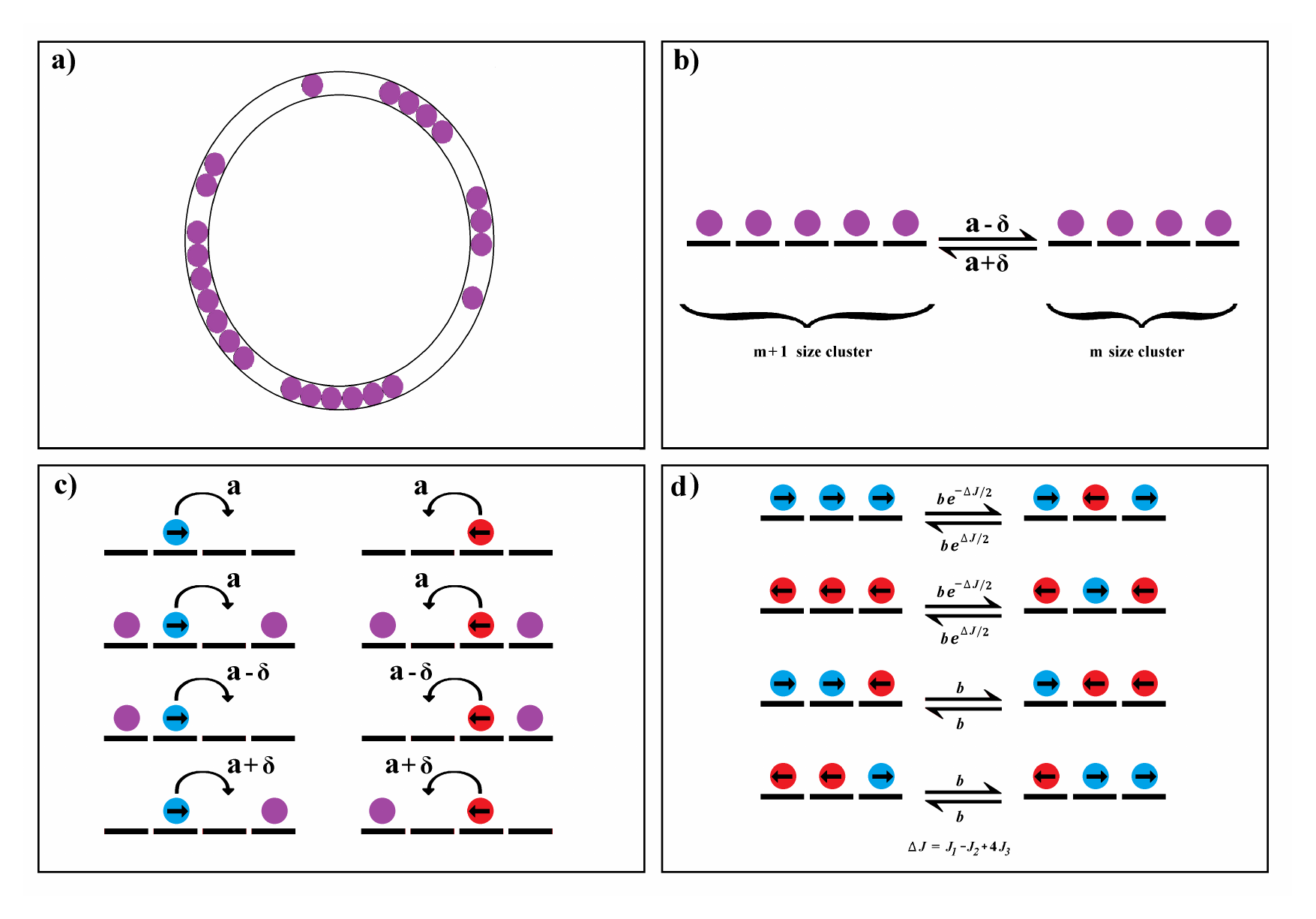}
	\caption{\textbf{Schematic representation of the system and dynamical processes:}  \textbf{(a)} Schematic of the cells confined within annular micropatterned substrate. \textbf{(b)} Schematic depicting the process of transformation of cluster comprising of $m+1$ particles to a cluster with $m$ particles. \textbf{(c)} Translation process of particles on the 1D lattice. \textbf{(d)} Process of polarity switching of a particle in the bulk of a cluster when they are bounded by other particles at both ends.}
	\label{fig:schematic}
\end{figure*} 

Theoretical investigation into the nature of cellular organization is based on the underlying governing paradigm that cellular systems are driven out-of-equilibrium systems, where the drive is at the level of individual cells \cite{sriram-rev,active-rev}. Different phenomenological modeling approaches, ranging from particle or agent-based models, active vertex model and, active hydrodynamics model have been used to gain insight into cellular organization and clustering properties of cellular systems \cite{agent1,agent2, igna-pnas, levine-pnas, levine-prl, soto, sriram-rev, hydro1, hydro2, ananyo,sm-scirep}. 
In this context, it may be noted that while diverse {\it continuum}  agent-based models \cite{agent1,agent2} and continuum hydrodynamic approaches \cite{ananyo,hydro1,hydro2} have been extensively used to elucidate the collective behavior of cells in tissues and cell colonies, the use of {\it discrete driven lattice gas} modeling approach \cite{schutz-rev,driven3} to study clustering characteristics in cell assemblies has remained rather sparse \cite{sm-scirep}. The scope and utility of adopting such a minimalist modeling approach can hardly be underscored. In fact, discrete driven lattice gas models have proven to be very useful in providing insight in a variety of physical and biological phenomena, such as  transport across bio-membranes \cite{chou}, intra-cellular transport \cite{sm-pre1, sm-pre2}, and growth process of fungal mycelium \cite{fungi1, fungi2, fungi3} among others. Recently discrete lattice gas modeling approach has been adopted to study the effect of CIL interaction on cell aseemblies in confined quasi-1D geometry \cite{sm-scirep}.

A widely used experimental technique to study {\it in-vitro} cell-cell interactions are the motility assays \cite{cil1,cil2,cil-plos,ananyo}. Recent advancement of this  experimental technique has led to the development of 1D collision assays, where cells motion is confined to micropatterned stripes of width slightly larger than typical size of the cells. Such 1D assays have proven to be very useful in ensuring the reproducibility of cell-cell collision event and quantifying the strength of the individual interactions between cells \cite{ananyo,sm-scirep,cil-plos}. 
Inspired by such experimental setups, we propose and study an active spin model in a discrete 1D lattice, to study the emergent properties of the cell assemblies arising out of the interplay of cell self-propulsion and generic cell-cell interactions. We focus our attention on the collective behavior in terms of collective spatio-temporal properties that can easily be amenable in such an experimental set-up. In particular, we investigate the characteristics of spatial organization by studying the cell clustering, e.g., cluster size distribution and average cluster size. 

\section{Model and Methods}
We propose a 1D discrete model that accounts for the phenomenon of cellular self-propulsion, their ability and propensity to switch their propulsion direction, the short range attractive and repulsive position dependent interactions between cells, and their polarity dependent interactions arising due to CIL other polarity alignment interactions.

We represent the 1D assay with annular geometry as a discrete 1D lattice comprising of $L$ sites while the $N$ cells  are considered as 'particles' (see Fig.\ref{fig:schematic} (a)). In this description, the individual cells are un-deformable objects of finite size, with the lattice spacing corresponding to the individual cell size. The microscopic state of the system at a given site $i$ is characterized in terms of occupation number variable $n_i$ and polarization vector $\sigma_i$ associated with the polarity (directionality of the particle movement). Since cell size prevents multiple occupancy, $n_i$ can only take values $1$ and $0$, depending on whether the site is occupied or empty. Moreover, $\sigma_i$ can take values $+1$ or $-1$, depending on whether the cell polarity (directionality of movement) is pointing towards the right or the left, respectively. The direction in which the particle hops is determined by its polarity state. While particle with $\sigma_i = +1$ can hop to the adjacent site $i+1$ if the site is empty, for particle with $\sigma_1 = -1$, it can hop to site $i-1$, provided the site is vacant. 

We define a multi-particle cluster as a continuous array of two or more particles,  bounded by vacancies at both ends. Due to short-range interaction, cells will not only exhibit an increased propensity to form multi-particle clusters, but will decrease their propensity to move out of a such multi-particle cluster. This effect is taken into account by choosing the translation dynamics of particles as prescribed in Fig. \ref{fig:schematic} (c). A single particle hops with a rate $a+\delta$ provided it forms a multi-particle cluster, whereas it hops with a rate $a-\delta$ if it comes out of the multi-particle cluster to a become a single particle. This dynamics would lead to a process of transformation between clusters of different sizes such that change of cluster of size $m+1$ to size $m$ maybe schematically represented as shown in Fig. \ref{fig:schematic} (b).

If this process satisfies detailed balance, the energy change associated with with a cluster of size $m$ changing to size $m+1$ can be expressed as, 
\begin{equation}
\Delta E_{a} =  -J_o = - \ln \left( \frac{a +\delta}{a -\delta} \right) 
\label{DE-KLS}
\end{equation}
Further, on account of this attractive interaction, for a cluster of size $m$, one can associate an energy of the form,
\begin{equation}
E_{a}(m) = -(m-1)J_o
\label{energy-Jo}
\end{equation}

Therefore, for a given configuration of particles, one can associate an effective interaction potential arising due to short-range attraction between cell, which can be expressed in the form, 
\begin{equation}
H_o =  -J_o\sum_{i} n_{i} n_{i+1}
\label{H-KLS}
\end{equation}

The defining feature of CIL interaction between cells is the tendency to realign their respective polarities away from each other. Such an interaction can be taken into account by considering an effective interaction potential between neighbouring particles which depends on their polarities of the form \cite{sm-scirep}, 
\begin{eqnarray}
H_{CIL}=\sum_{i} n_{i} n_{i+1}\left[ J_1 \Theta( \sigma_{i} - \sigma_{i+1} ) - J_2\Theta( \sigma_{i+1} - \sigma_{i} ) \right],\nonumber
\label{eqn-H}
\end{eqnarray}
where $\Theta$ is the Heaviside function. When  $J_1, J_2 >0$,  configurations for which neighbouring particles point to each other are disfavoured, while configurations with oppositely aligned polarity of neighbouring particles is favoured. 

The propensity of cells to favour alignment of their polarities can be expressed as an effective Ising ferromagnetic like interaction term,
\begin{equation}
H_f = -J_3\sum_{i} n_{i} n_{i+1}(\sigma_{i}\sigma_{i+1}) \nonumber
\end{equation}
with $J_3 >0$. 

Thus, overall, the interaction potential arising due to these two different alignment interactions and short-range attractive interaction between particles is expressed as,

\begin{eqnarray}
H_T = \sum_{i} &n_{i} n_{i+1}\left[ J_1 \Theta( \sigma_{i} - \sigma_{i+1} ) - J_2\Theta( \sigma_{i+1} - \sigma_{i} )\right.\nonumber \\
 &\left. - J_3 \sigma_{i}\sigma_{i+1} - J_o\right]
 \label{eq:interaction}
\end{eqnarray}

Particle switching dynamics between different polarization states is governed by the interaction potential in Eq. (\ref{eq:interaction}). We set $k_b T =1$, where temperature $T$ is an effective temperature for particles arising due to activity. Then the particle polarity switching rate, $k_o \propto exp(\Delta E)$, where $\Delta E$ is the energy difference between the two microscopic configurations. A schematic representation of the relevant dynamical processes associated with polarity switching of particles, is given in Fig.~\ref{fig:schematic}(d). The switching and translation dynamics rates are summarized in Appendix \ref{app:simulation-Details}.

\subsection{Connection and mapping to other models}

In the absence of CIL interaction, alignment interaction and attractive interaction between particles (which corresponds to setting $J_1, J_2, J_3$ and $J_o$ to zero), the dynamics of this model is identical to Persistent Exclusion process (PEP) in 1D \cite{soto}. It may further be noted that in the absence of switching dynamics and particles with same polarity, the model further reduces to nearest neighbour Totally Asymmetric Exclusion Process (TASEP). However, when the attractive interaction is present, i.e., when $J_o\neq 0$, the dynamics of the particles at a given site also depends on the next nearest site. In this scenario, the model corresponds to a Next-Nearest Neighbour (NNN-TASEP). It is also identical to the Katz-Lebowitz Spohn (KLS) model \cite{KLS-main, KLS-1, KLS-2} and the corresponding Hamiltonian for the system can be expressed as,
\begin{equation}
H_{KLS} =  -J_o\sum_{i} n_{i} n_{i+1}
\end{equation}
where, $J_o = \ln \left( \frac{a +\delta}{a -\delta} \right) $ and $n_i$ are occupation numbers at site $i$ which can take values 0 or 1. 

When short range inter-particle attraction is absent ($J_o = 0$) while CIL and other alignment interactions are present ($J_1,J_2, J_3 \neq 0$) along with polarity switching dynamics of particles ($b \neq 0)$, the model reduces to exactly the same form described in Ref.\cite{sm-scirep}. 
In this regime, Eq. (\ref{eqn-H}) can be reorganized into an Ising-like alignment term and a symmetric CIL-like interaction term, leading to

\begin{eqnarray}
H  = \sum_{i}& -n_i n_{i+1}\left[ \left(\frac{J_1 - J_2 + J_3}{4}\right) {\bf \sigma}_{i}\cdot{\bf \sigma}_{i+1}\right. \nonumber \\ &\left. - \left(\frac{J_1 + J_2}{4}\right) ({\bf \sigma}_{i} - {\bf \sigma}_{i+1})\cdot {\bf \hat r}_{i,i+1}\right],
\label{eqn-H1}
\end{eqnarray}
where, ${\bf \hat r}_{i,i+1}$ is a unit vector along ${\bf r_i}-{\bf r_{i+1}}$. In this expression, the first term corresponds to the Ising-like interaction term, and second term can be interpreted as an interaction term associated with repulsion due to CIL. This term features only as a boundary term and it does not contribute to the energy in the bulk of the cluster. The second term of Eq.\ref{eqn-H1} is similar in form to that of interaction potential arising from the coupling between polar order and nematic splay in the context of nematic liquid crystals \cite{nematic}. 
When $J_3 =0$, for the symmetric case i.e., when $J_1 = J_2$, the Ising-like contribution cancels and the interaction term associated with repulsion due to CIL does not contribute to the energy in the bulk.
However, in the presence of vacancies, this interaction term plays a crucial role in the cluster organization. 

\section{Results}
\subsection{Fully packed state: a reduced equilibrium model}
\begin{figure*}
	\centering
	\includegraphics[width = \linewidth]{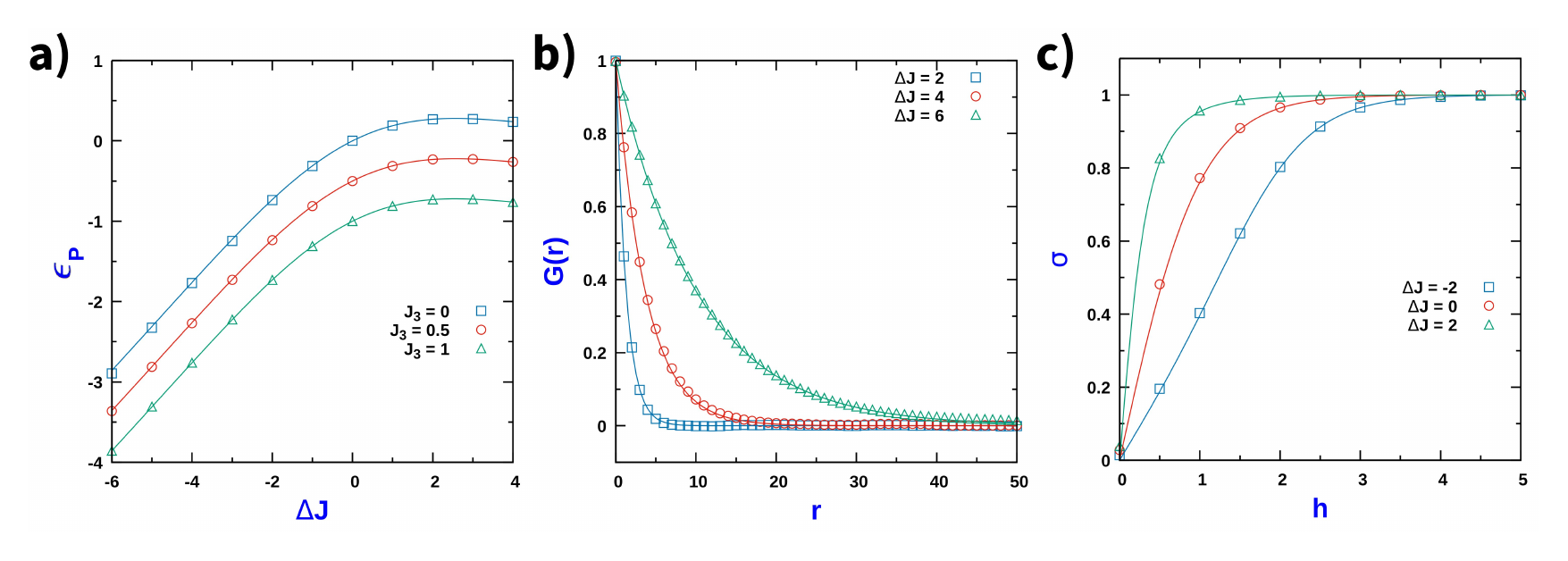}
	\caption{\textbf{(a)} Variation of the average energy per particle $\epsilon_{p}$ as a function of $\Delta J$ for different values of $J_3$. $\epsilon_{p}$ is expressed in units of $k_B T$, with $T$ being the effective active temperature. Solid lines correspond to the expression in Eq. (\ref{eqn-avgE-CIL}). \textbf{(b)} Correlation function for polarization $G(r)$ as function of particle distance $r$ for different $\Delta J$. Solid lines correspond to the expression in Eq. (\ref{eqn-corr-CIL}). \textbf{(c)} Variation of average polarization per particle $\sigma$ as a function of external field $h$ for different $\Delta J$. Solid lines correspond to Eq. (\ref{eqn-avgMag-CIL}). For all cases, the points corresponds to results obtained using MC simulations. MC simulation was done with $L=2000$ and averaging was done over $2000$ samples.}
	\label{fig:fully_packed}
\end{figure*}

{\it In-vitro} experiments with {\it Madin-Darby Canine Kidney}  (MDCK) cells in quasi-1D substrate have been performed with different cell number densities \cite{ananyo}. At high densities it has been observed that the cells attain a confluent state and  cover the entire 1D substrate \cite{ananyo}. From the standpoint of our Active spin model, this regime corresponds to a fully packed state where the particle translation motion is arrested and the particles are only allowed switch their polarity. In this regime, the system behaves as an effective equilibrium spin system.
Invoking the Transfer Matrix method technique, we obtain the expression for the average energy per particle, $\epsilon_{p}$ in the thermodynamic limit ( $N\rightarrow \infty$) as, 
\begin{equation}
\epsilon_{p} = -J_{3} + \frac{\Delta J}{2}\left[\frac{1}{1 + \exp \frac{\Delta J}{2}}  \right ] 
\label{eqn-avgE-CIL}
\end{equation}
Where, $\Delta J = J_1 - J_2 + 4J_3$. While the average polarization is zero, the correlation function for the polarization as a function of particle distance $r$, assumes the form,
\begin{equation}
G(r) ={\left( \frac{1 - \exp(-\frac{\Delta J}{2})}{1 + \exp(-\frac{\Delta J}{2}) }\right )}^{r}
\label{eqn-corr-CIL}
\end{equation}

In the presence of an external field, $h$, which couples with the individual particle polarization, the expression for the average polarization per particle $\sigma$ assumes the form, 
\begin{equation}
\sigma = \frac{\sinh(h)}{\left[ \cosh^{2}(h) + \exp (-\Delta J) -1\right]^{1/2}}
\label{eqn-avgMag-CIL}
\end{equation}

Fig.~\ref{fig:fully_packed} shows the comparison of the analytical expressions with MC simulations.

\subsection{Clustering in the absence of CIL and alignment interaction}

\subsubsection{Reduction to KLS model}
We now focus on the regime where the polarization alignment interactions are weak or negligible, which corresponds to $J_1 = J_2 = J_3=0$. If the rate of polarity switching, $b$, is much faster than the rate of hopping $a$, corresponding to the limit, $Q \ll 1$,  the model can be approximated by an effective KLS model. Accordingly, we use the fact that detailed balance is satisfied implying that the relevant rate is a function of the energy change, $\Delta E$, when a cluster changes from size $m$ to $m+1$, and that is a function of $a$ and $\delta$, as in Eq. (\ref{DE-KLS}). Further it implies that for this case, the dynamics of the system is determined  by an effective Hamiltonian that coincides with the KLS Hamiltonian, 
\begin{equation}
H_T = -J_o\sum_{i} n_{i} n_{i+1}
\end{equation}
where $J_o = \ln\left(\frac{a+\delta}{a -\delta}\right)$ and $n_i$ are the occupation number variables that can take value $1$ or $0$.

The KLS model can be mapped on exactly to an effective Ising model and the corresponding Hamiltonian can be expressed as
\begin{equation}
H_{T}=-\frac{J_o}{4}\sum_{i}s_{i} s_{i+1} 
\end{equation} 
where, $s_{i} = \pm 1$. The Ising spin variables ${s_i}$ are related to ${n_i}$ by the relation, $s_i = 2n_i -1 $. The fact that the number density is held fixed implies that,  in the language of Ising spin system, this system is in a constant 'magnetization' ensemble, where the effective constant magnetization $M$, in terms of the density of particles in lattice $\rho$ assumes a form, $M = \sum_{i}s_i=2L\left(\rho-\frac{1}{2}\right)$. 

\subsubsection{Partition function and Average Energy}

The corresponding Gibb's partition function in the constant density ensemble reads, 

\begin{equation}
Z_G(L,\rho)=\sum_{s_i}\exp{\left[\sum_{i}\left( \frac{J_o}{4} s_{i} s_{i+1}+ h_os_i \right)\right]}
\label{Zg1}
\end{equation}

Using Transfer Matrix method, in the thermodynamic limit of $L\rightarrow \infty$, we arrive at
\begin{equation}
Z_G(L,\rho)=\left\{e^{\frac{J_o}{4}} \left[\cosh{(h_o)} + \sqrt{\sinh^{2}{(h_o)}+e^{-J_o}}\right]\right\}^{L} 
\end{equation}

The corresponding expression for the field $h_o$ is,
\begin{equation}
h_o= \sinh^{-1}\left[\frac{\left(\rho -\frac{1}{2}\right) e^{-\frac{J_o}{2}}}{\sqrt{\rho(1-\rho)}}\right]  
\label{eqn-h}    
\end{equation}

Using the expression of $h_0$, the general expression for the Gibb's Partition function reads as, 

\begin{eqnarray}
&Z_G&(L,\rho)=\left[ \frac{e^{\frac{J_o}{4}}}{\sqrt{\rho(1-\rho)}}\right]^{L} \times  \nonumber\\ 
&&\left[ \left( \frac{e^{\frac{-J_o}{2}}}{2}+\sqrt{\rho(1-\rho)+ \left( \rho-\frac{1}{2} \right)^{2}e^{-J_o}} \right) \right]^{L} 
\label{Zg}  
\end{eqnarray}

The corresponding expression for the average energy of the system reads as,

\begin{equation}
\left \langle E \right \rangle_A=-\frac{J_oL}{4}\tanh\left ( \frac{J_o}{4} \right )-J_oL\left ( \rho-\frac{1}{4} \right )= L\epsilon_A
\label{eqn-avgE-KLS}
\end{equation}

\begin{figure*}
	\centering
	\includegraphics[width = \linewidth]{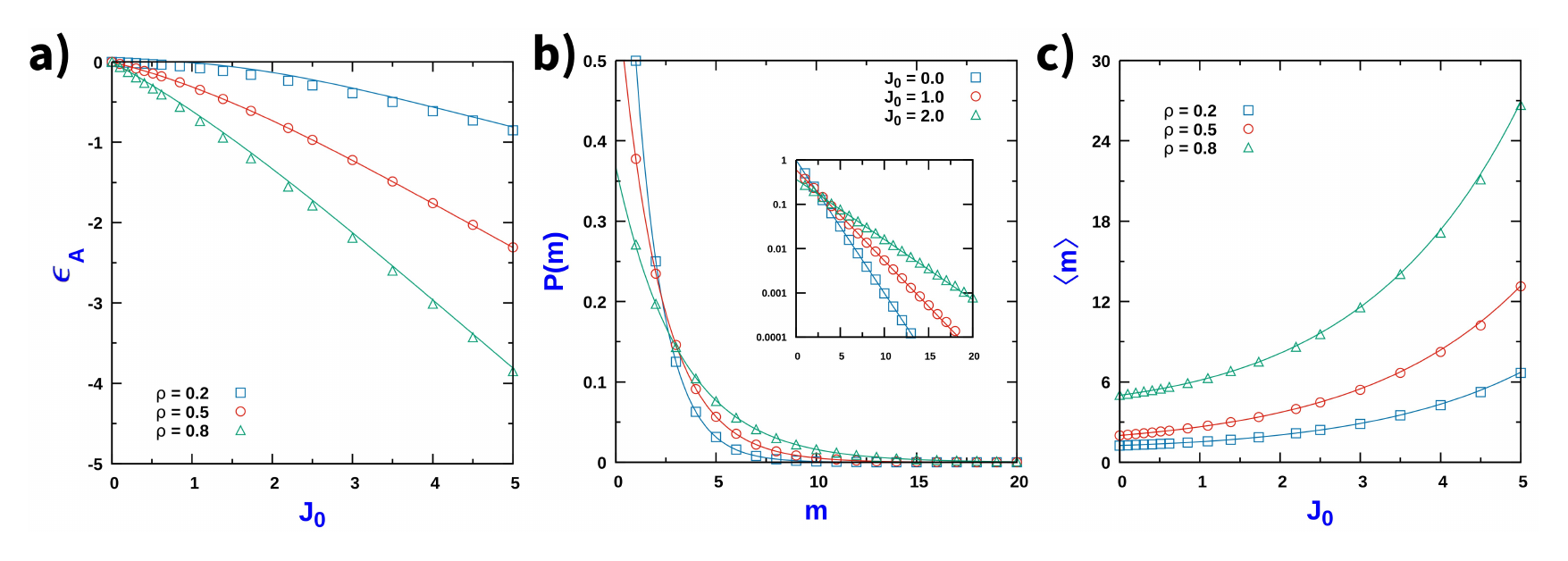}
	\caption{\textbf{(a)} Variation of average attractive energy per site $\epsilon_A=\frac{\left \langle E \right \rangle_A}{L}$ as function of $J_o$. Here (i) $\rho = 0.2$, (ii) $\rho = 0.5$, (iii) $\rho = 0.8$. The solid lines correspond to the expression in Eq. (\ref{eqn-avgE-KLS}). \textbf{(b)} Corresponding Cluster size distributions $P(m)$ for different $J_o$ for $\rho = 0.5$. The solid lines in correspond to the expression in Eq. (\ref{eqn-CSD-KLS}). Inset figure is the corresponding semi-log plot. \textbf{(c)} Corresponding average cluster size $\left \langle m \right \rangle$ as function of $J_o$. The solid lines correspond to the expression in Eq. (\ref{eqn-avgCS-KLS}). For all cases, points correspond to MC simulations. For all cases $J_1 = J_2 = J_3 = 0$, $a=0.01$, $b=1$ $(Q = 0.01)$. MC simulation was done with $L=2000$ and averaging was done over $5000$ samples.}
    \label{fig:KLS}
\end{figure*}

Comparison of the analytical result of the average energy per site obtained from Eq. (\ref{eqn-avgE-KLS}) with MC simulations shows a very good agreement for a wide range of densities and $J_o$ in the limit of $Q \ll 1$ ( See Fig. \ref{fig:KLS} (a)). 

\subsubsection{Cluster size characteristics}
A cluster of size $m$ is a continuous stretch of $m$ particles, with vacancies on both sides. A general configuration of a cluster of size $m$ can be schematically represented as,
\begin{equation}
-1\hspace{0.1cm}\underbrace{11111...1}_{m}\hspace{0.1cm}-1  \nonumber  
\end{equation}
The probability of a cluster of size $m$ with $L$ sites i.e. $P_{L,\rho}(m)$ can be expressed in terms of the probability of finding a cluster of m occupied sites (particles) with vacancies on both sides of the cluster i.e. $p(m)$ as,

\begin{equation}
P_{L,\rho}(m)=L \times  p(m) \nonumber  
\end{equation}

\begin{equation}
p(m)=\frac{\underbrace{e^{\frac{J_o}{4}(m-3)+ h_0(m-2)}}_{(a)}\times \underbrace{Z_{G}^{*}(L-m-2, \rho_{*})}_{(b)}}{Z_{G}(L, \rho)} \nonumber 
\end{equation}

This form, (a) corresponds to the Boltzmann weight associated with cluster of size $m$, (b) corresponds to the Gibbs partition function for the remaining 
$L-m-2$ sites with number density $\rho_{*}=\left( \frac{N-m}{L-m-2} \right)$. 
The Gibbs partition function in Eq. (\ref{Zg}) can be written as $Z_G = K^{L}$ where, 
$ K=\left[ \frac{e^{\frac{J_o}{4}}}{\sqrt{\rho(1-\rho)}}\right] \times  \left[ \left( \frac{e^{\frac{-J_o}{2}}}{2}+\sqrt{\rho(1-\rho)+   \left( \rho-\frac{1}{2} \right)^{2}e^{-J_o}} \right) \right]$.
In the thermodynamic limit, $L \to \infty$, $Z_G^{*} \simeq K^{L-m-2}$ and $P_{\rho}(m)$ can be expressed as,
$P_{L,\rho}(m)\propto e^{m\left( \frac{J_o}{4} +h_0\right)}K^{-m}$. Using the normalization condition it follows that the normalized cluster size distribution function $P(m)$ for a fixed number density $\rho$ in the lattice is,
\begin{equation}
P(m)=\left( e^{D}-1 \right)e^{-Dm}
\label{eqn-CSD-KLS}
\end{equation}
where, 

\begin{eqnarray}
D&=&-\sinh^{-1} \left[ \frac{\left( \rho-\frac{1}{2} \right) e^{-\frac{J_o}{2}}}{\sqrt{\rho(1-\rho)}} \right]\\&+& \ln\left[ \frac{e^{-\frac{J_o}{2}}}{2 \sqrt{\rho(1-\rho)}}+\frac{\sqrt{\rho(1-\rho)+\left( \rho-\frac{1}{2} \right)^{2}e^{-J_o}}}{\sqrt{\rho(1-\rho)}}\right] \nonumber
\label{eq-D}
\end{eqnarray}
Using Eq.(~\ref{eqn-CSD-KLS}), the average cluster size reads as,
\begin{equation}
\left\langle m \right\rangle=\sum_{m=1}^{\infty}mP(m) = \frac{e^D}{e^D -1}
\label{eqn-avgCS-KLS}
\end{equation}
In the limit of $J_o = 0$, the model reduces to SEP and, as expected, the distribution function of the cluster size $P(m)$ coincides with the form of PDF for SEP, with the explicit expression being,
\begin{equation}
P(m) = \left(\frac{1-\rho}{\rho}\right) e^{-m/\xi}
\end{equation}
where, $\xi = {| \ln \rho |}^{-1}$.

When $J_o \neq 0$, and $\rho = 1/2$, the expression for $P(m)$ simplifies to 
\begin{equation}
P(m)=e^{-\frac{J_o}{2}}e^{-D m}
\end{equation}
where $D = \ln~(1 + e^{-\frac{J_o}{2}})$.
The corresponding expression for average cluster size is,
\begin{equation}
\langle m \rangle = 1 + e^{\frac{J_o}{2}}
\end{equation}
In this limit, the effect of the attractive interaction between particles is to increase the cluster size compared to the non-interacting case. 

Fig. \ref{fig:KLS}(b) shows comparison of the analytic expression of $P(m)$ , Eq. (\ref{eqn-CSD-KLS}) with MC simulation. Fig. \ref{fig:KLS} (c) shows the comparison of the analytic expression for $\left \langle m \right \rangle$, Eq. (\ref{eqn-avgCS-KLS}), with MC simulation results. 

In the limit of $ Q \ll 1$, excellent agreement of the analytical expressions for average energy of cluster, cluster size distribution and average cluster size with MC simulation validates the mapping to the KLS model that we have proposed. It may however be noted that away from this limit, the mapping to KLS model no longer holds good ( See Fig. \ref{fig:KLS-highQ} in Appendix \ref{app:additional-figures}).

\subsection{Clustering in the presence of CIL and alignment interaction}

Clustering arises when a motile particle encounters another particle due to the excluded volume. Conversely, the disintegration of the cluster occurs due to the polarity switching of particles at the cluster boundaries. Thus, whenever the typical collision rate $a\rho$ is larger than the switching rate $b$, clusters will emerge.
Consequently, the composition of the cluster is characterized by an array of right-pointing $(\rightarrow)$ particles occupying the left end of the cluster, while the right end comprises an array of left-pointing $(\leftarrow)$ particles. The internal structure within the bulk of such clusters consists of defect—pairs of $(\rightarrow)$ and $(\leftarrow)$ particles. It is also noteworthy that clusters with particles at both ends pointing inward are immobile. At any instant, a cluster size can increase if a $(\rightarrow)$($(\leftarrow)$) particle from the adjoining gas region joins the left(right) end of the cluster. A cluster size can decrease if a single particle at the cluster boundaries switches its polarity and subsequently leaves the cluster.

Fig. ~\ref{fig:morphology} (a) in Appendix \ref{app:additional-figures} displays the spatio-temporal evolution of the clusters for different CIL strength as a function of $J_1$ for $Q \gg 1$. The average cluster sizes  increases with $J_1$. Similarly, Fig. ~\ref{fig:morphology} (b) displays the spatio-temporal evolution of the clusters for different attractive interaction as a function of $\delta$ in the $Q \gg 1$ regime. For $\delta = 0$ the system segregates into alternating domains of dense clusters and a low-density gas region. As we increase $\delta$ and in the $a \sim \delta$ regime, single particles in the gas region are absorbed into dense clusters, leading to the formation of stable dense clusters.

\subsubsection{Universality of cluster size distribution when $Q\gg 1$}
When the translation rate is larger than the polarity switching rate, $Q \gg 1$, the system segregates into alternate regions of dense clusters(c) phase and a low density gas phase(g). It maybe noted that for the dense cluster(c) phase, the smallest cluster comprises of at least two particles. In this limit of $Q \gg 1$, the interaction between the dense clusters is weak and the stationary state is achieved as a balance between the incoming particle flux into the dense clusters from the surrounding gas region and the outgoing particle flux from the dense clusters due to the polarity switching at the cluster ends~\cite{soto}. 

One can define configurational entropy $\mathcal{S}$ of the dense cluster phase as the number of ways of arranging $\Omega$ clusters in a manner that clusters of same size are indistinguishable, when there is a constraint that the total number of sites occupied by the clusters, $N_c$, and $\Omega$ is held fixed. In the absence of any interactions, the cluster size distribution is obtained by minimizing the system configurational entropy, as has been shown for Persistent Exclusion Process (PEP)~\cite{soto}. In the presence of inter-particle interactions, the system can be mapped to an equivalent equilibrium system where the cluster size distribution is determined by minimizing an effective Helmholtz free energy which includes the energy contributions due to interactions \cite{sm-scirep}.  In our case,  attractive interactions between particles, CIL and alignment interactions between particle polarities are  accounted for through an additional internal energy contributions whose general forms are provided in Eq. (\ref{energy-Jo}) and Eq. (\ref{eqn-avgE-CIL}) (in the limit of large cluster size). Thus the general form of the effective Helmholtz free energy functional, $\mathcal{F}$ is,
\begin{eqnarray}
\mathcal{F}&=&- \ln \left ( \frac{\Omega !}{\prod_{m} G_{c}(m)!} \right )
+ \sum_{m} \left [E_p(m) + E_a(m)\right]G_{c}(m) \nonumber\\
&+& \lambda \left ( N_{c} - \sum_{m} m G_{c}(m)\right ) + \gamma \left (\Omega -\sum_{m} G_{c}(m) \right )  
\label{eqn-F}
\end{eqnarray}
where, $G_{c}(m)$ is the number of clusters of size $m$ in the dense cluster phase. The first term in Eq. (\ref{eqn-F}) corresponds to the configurational entropy, while the second and third terms account for the energy of a cluster of size $m$ due polarity and attraction, respectively. $\lambda$  and $\gamma$ are the Lagrange multipliers to ensure the constraints on $N_c$ and $\Omega$, respectively.
$E_a(m)= -( m - 1 )J_o$, and  the explicit general form for $E_p(m)$ is given in  Appendix \ref{app:calculations-ae}. 

The steady state cluster size distribution in dense cluster phase, $P_c(m)$, is determined by minimizing $\mathcal{F}$ (see Appendix \ref{app:calculations-acs} for further details)

\begin{figure}[t!]
\centering
\includegraphics[width = \linewidth, height = 5cm]{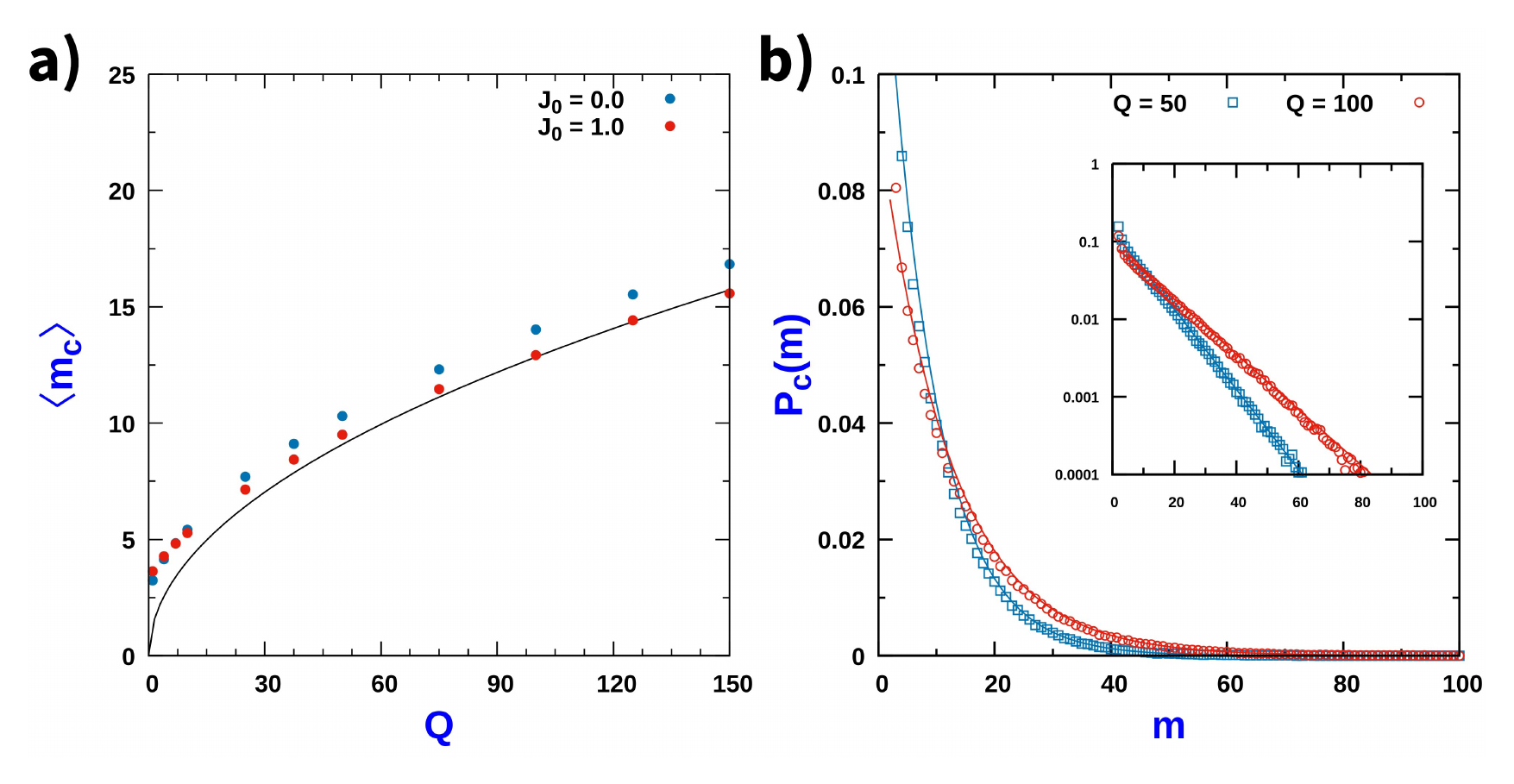}
\caption{\textbf{(a)} Variation of average cluster size in the dense cluster phase $\left \langle m_{c} \right \rangle$ as a function of $Q$ for different values of $J_o$. Solid lines corresponds to analytical expression in Eq. (\ref{mc-form}) while circles corresponds to MC simulations. \textbf{(b)} Cluster size distribution in dense cluster phase, $P_c(m)$ as a function of $m$ for different $Q$. Solid lines corresponds to analytical expression in Eq. (\ref{eqn-CSD_dense}) while circles corresponds to MC simulations. For all cases particle number density $\rho = 0.5$ , $J_{1} = 1$, $J_{2} = J_{3} = 0$, $b = 1$. MC simulation was done with $L = 2000$ and averaging was done over $5000$ samples.}
\label{fig:compare2}
\end{figure}

\begin{figure*}
	\centering
	\includegraphics[width =\linewidth]{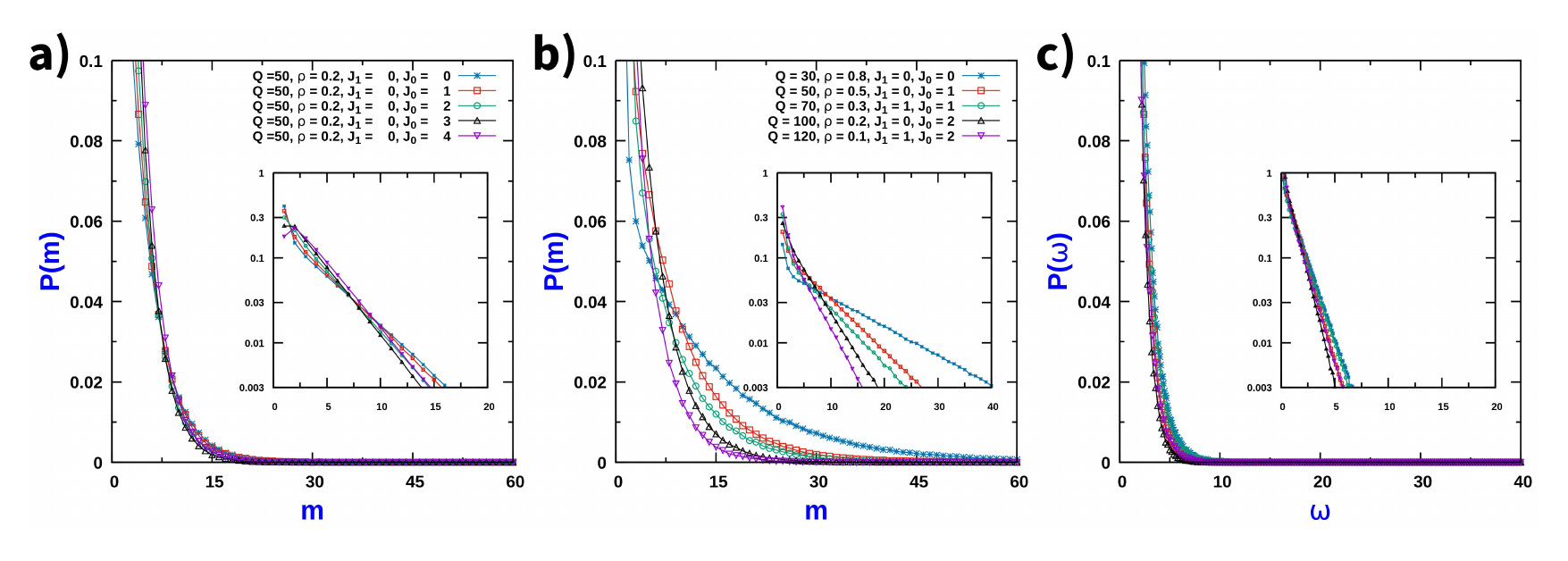}
	\caption{ Cluster size distribution in the $Q \gg1$ regime: \textbf{(a)} $P(m)$ vs $m$ for different $J_o$. Here, $Q=50$, $\rho =0.2$, and $J_1 =0$. \textbf{(b)} $P(m)$ vs $m$ for different $Q$, $\rho$, $J_1$ and $J_o$. \textbf{(c)} $P(\omega)$ vs $\omega$, where the scaled variable $\omega=m\left ( \frac{1-\rho}{Q\rho} \right )^{1/2}e^{-J_1/4}$, for the same data set as in (b). Inset figures are the corresponding log-log plots. For all cases $J_2=0, J_3=0$, $b=1$. MC simulation was done with $L=2000$ and averaging was done over $5000$ samples.}
    \label{fig:data_collapse}
\end{figure*}

Minimizing the free energy functional, $\delta \mathcal{F}  = 0$, we can derive form of the normalized cluster size distribution in dense cluster phase, $P_c(m)$ (comprising of clusters of size 2 and above).
The details of the derivation steps are discussed in Appendix \ref{app:calculations-ae}.  
In the thermodynamic limit of $L \rightarrow \infty$, we get, 
\begin{equation}
P_{c}(m)=A_{c}e^{-\frac{m}{m_{c}}} 
\label{eqn-Gc2_main}
\end{equation}

A similar argument for the gas phase, yields an exponential form of $P_{g}(m)$, 
\begin{equation}
P_{g}(m)=A_{g}e^{-\frac{m}{m_{g}}} 
\label{eqn-Gc4_main}
\end{equation}

The parameters $A_{c}$, $A_{g}$, $m_{c}$ and $m_{g}$ that identify the coexisting cluster size distributions of the gas and dense cluster are determined using appropriate boundary and continuity conditions, leading to the corresponding general form for the average cluster size ( For details see Appendix \ref{app:calculations-acs}).

As long as $Q \gg e^{J_o}, e^{J_1}$, the expression for the average cluster size in the cluster phase reduces to,

\begin{equation}
\langle m_c \rangle = {\left( \frac{\rho}{1-\rho}\right)}^{1/2} e^{\frac{J_1}{4}}~Q^{1/2} + \mathcal{O}(Q^{0}) + \mathcal{O}(Q^{-1/2})
\label{mc-form}
\end{equation}
It follows that up to leading order in power of Q, $\langle m_c \rangle \propto Q^{1/2}$ and that the average cluster size does not not depend on the attraction strength $J_o$ (this  contribution is subdominant in powers of $Q$). The expression of $\langle m_c \rangle$ is exactly the same as obtained for the case discussed in Ref.{\cite{sm-scirep}}, where 1D model with only CIL interaction was studied. 
The explicit form of the normalized PDF of the cluster sizes in the dense phase reads as, 
\begin{equation}
    P_c(m) =\left ( e^{\frac{2}{\xi}}- e^{\frac{1}{\xi}}  \right ) e^{- \frac{m}{\xi}}
    \label{eqn-CSD_dense}
\end{equation}
where,
$1/\xi=-\ln\left( \frac{\left\langle m_c \right\rangle-2}{\left\langle m_c \right\rangle-1} \right)$ and where the corresponding expression of  $\langle m_c \rangle$ is given by Eq. (\ref{mc-form}). 
In Fig. \ref{fig:compare2} (a) we display the comparison of  the analytical expression of the average cluster size in the dense cluster phase $\left \langle m_{c} \right \rangle$ in Eq.~(\ref{mc-form}) with  MC simulations. Fig. \ref{fig:compare2} (b) displays the analogous comparison for $P_c(m)$. In the limit $Q \gg e^{J_o}, e^{J_1}$, the cluster size distribution approaches an exponential form of Eq. (\ref{eqn-CSD_dense}) with $\langle m \rangle$ given by Eq. (\ref{mc-form}). 

Fig. \ref{fig:data_collapse} displays $P_c(m)$ for different  sets of $J_o$,  $Q$ and $\rho$, while Fig. \ref{fig:data_collapse}(c) shows the collapse of  $P_c(m)$ into a universal curve when the cluster size is scaled into the variable $\omega=m\left ( \frac{1-\rho}{Q\rho} \right )^{1/2}e^{-J_1/4}$.
For moderate  CIL strength,  the scaling behaviour is fairly robust. The form of average cluster size also implies that in general in the limit of $Q \gg 1$, typical cluster size is controlled very strongly ( exponentially) by the CIL interaction strength. 

\begin{figure*}[t!]
    \centering
    \includegraphics[height = 10 cm, width = 0.9\linewidth]{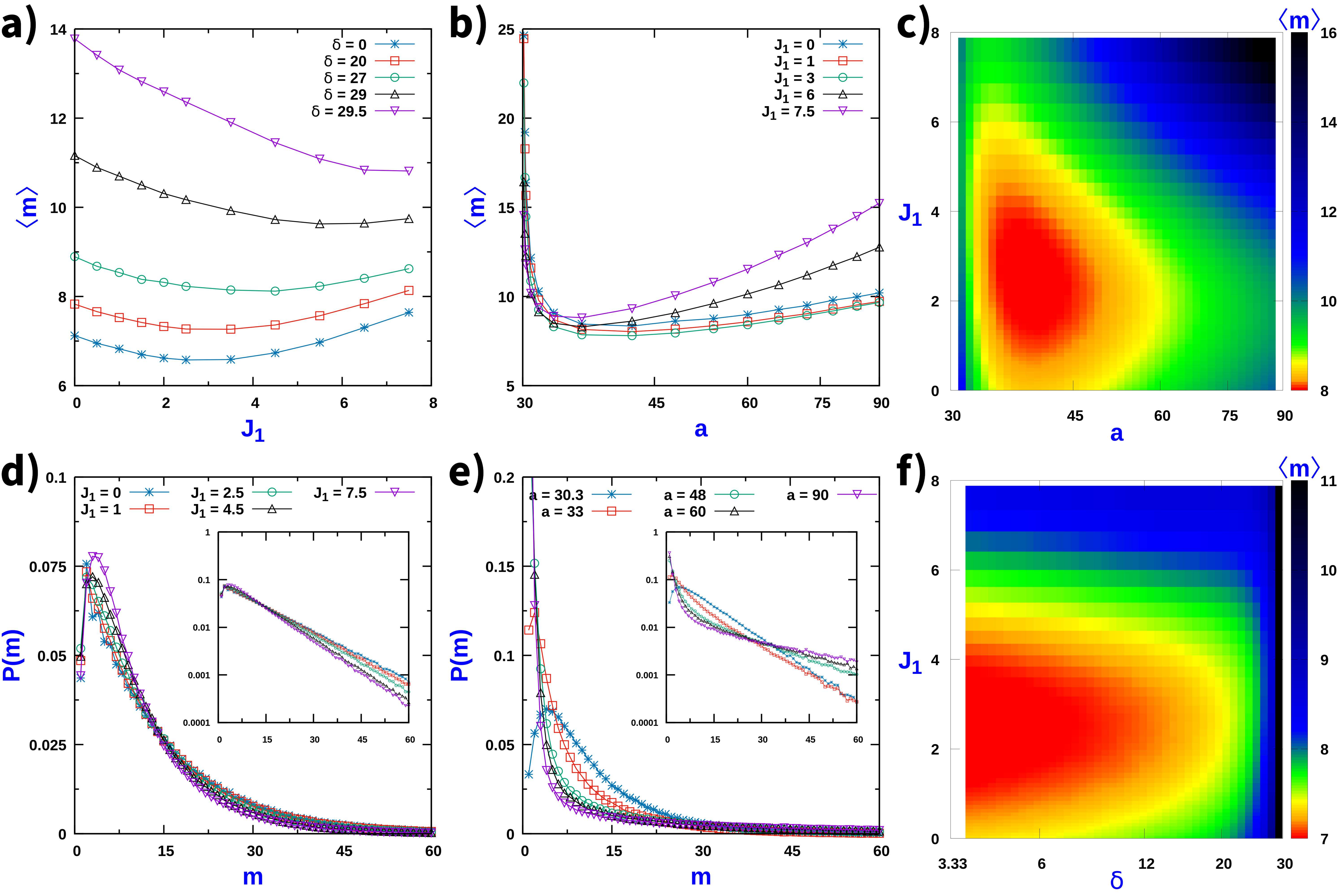}
    \caption{\textbf{Cluster size characteristics}: 
    \textbf{(a)} Average cluster size $\left \langle m \right \rangle$ as function of $J_1$ for different $\delta$ when $Q \gg 1$  $(a=30)$. \textbf{(b)} Average cluster size $\left \langle m \right \rangle$ as function of $a$ for different $J_1$ when $\delta^{'} \gg 1$ $(\delta = 30)$. \textbf{(c)} Corresponding Contour plot of average cluster size $\left \langle m \right \rangle$ as a function of $J_1$ and $a$ for $\delta = 30$.  \textbf{(d)} $P(m)$ vs $m$ for different $J_1$ for $\delta = 29.5$. Here $a = 30$. \textbf{(e)} Cluster size distributions $P(m)$ vs $m$ for different $a$ for $J_1 = 7.5$.Inset figures in (d) and (e) are the corresponding semi-log plots. \textbf{(f)} Contour plot of average cluster size $\left \langle m \right \rangle$ as a function of $J_1$ and $\delta$ when in high $Q$ regime $(a=30)$. For all cases, $\rho = 0.8$, $J_2 = 4$, $J_3 = 0$, $b = 1$ and MC simulation was done with $L=2000$ and averaging was done over $2000$ samples.} 
    \label{fig:cs}
\end{figure*}

\subsubsection{General cluster size properties}
We focus on the nature of clustering in cell assemblies due to the interplay of CIL strength through $J_1$ and $J_2$, the aligning interaction strength $J_3$, the translation hopping rate $a$, and the attractive interaction strength $\delta$. Cluster size is controlled by the interplay of $Q=\frac{a}{b}$ (ratio of transition rate $a$ to switching rate $b$ which is basically a measure of particle activity), $\delta^{'}=\frac{\delta}{b}$ (ratio of attractive interaction strength $\delta$ to switching rate $b$ which is an effective particle attraction) and effective CIL strengths $J^{'}_{1}=J_1+2J_3$, $J^{'}_{2}=J_2-2J_3$. Particle number density, $\rho$, also plays a significant role in the cluster size. In the following sections, we will systematically study the effect of the different parameters. For convenience, we set $b=1$ without loss of generality, which implies $Q=a$ and $\delta^{'}=\delta$.  

When $Q-\delta^{'} \gg 1$ the system segregates into alternating domains of dense clusters and a low-density gas. Particles displace larger distances before polarity switching  becomes effective, leading to the formation of large clusters. The average cluster sizes  increase with $Q$. When $Q-\delta^{'} \ll 1$, particles at cluster boundaries take much longer to escape, promoting an increase of the average cluster size when $a \simeq \delta$. 

 CIL interaction strength is characterized in terms of $J_1$ and $J_2$. While $J_1$ represents the energy cost associated with the polarities of neighboring particles pointing towards each other, $J_2$ represents the energy reduction associated with the polarities of neighboring particles pointing away from each other. In the absence of CIL interaction in a cluster, the switching rate from $(\rightarrow)$ to $(\leftarrow)$ is identical to the switching rate from $(\leftarrow)$ to $(\rightarrow)$. The interaction term of the Hamiltonian associated with CIL breaks the symmetry associated with independent rotation of the polarization vector. 

 In order to understand the role of CIL interaction on cluster size characteristics, we examine the role of the CIL parameters $J_1$ and $J_2$ individually. It maybe noted that while $J_1$ favours alignment of particle polarity, $J_2$ favours anti-alignment. At the cluster edges, while $J_2$ enhances the propensity of the particles to flip outwards, the effect of $J_1$ is just the opposite.
Hence $J_1$ promotes favours cluster size growth while $J_2$ has the opposite effect. If $J_1 \neq J_2$, which corresponds to an asymmetric CIL interaction, the effect of these parameters on the cluster size characteristics is more involved. The interplay of these two competing effects due to $J_1$ and $J_2$ manifests as a non-monotonic dependence of the average cluster size on $J_1$, for a fixed value of $J_2$, beyond a threshold value of $Q$ \cite{sm-scirep}. For experiments performed for cell aggregates in microfabricated quasi-1D  substrate, it has been observed that  typically for any single cell which is not in contact with other cell,  no polarity switching has been observed \cite{ananyo}. So we focus now on a situation for which polarity switching of a single particle which is not part of the cluster is explicitly disallowed. 

In the $Q \gg 1$ regime $(Q = 30)$, in absence of attractive interaction $(\delta=0)$, we can observe non-monotonous re-entrant-like behavior of average cluster size $\left \langle m \right \rangle$ due to the competition between  $J_1$ and $J_2$. As we increase $\delta$, this non-monotonic behavior persists ( See Fig. \ref{fig:cs}(a)). In the $\delta^{'} \gg 1$ region ($\delta^{'}$ = 30), we also observe a non-monotonous re-entrant-like behavior in the average cluster size $\left \langle m \right \rangle$ as a function of hopping rate, $a$ due to the competition between the effects of $a$ and $\delta$ (see Fig.\ref{fig:cs} (b)). Fig. \ref{fig:cs} (d) and Fig. \ref{fig:cs} (e) shows the variation of $P(m)$ with CIL strength parameter $J_1$, and particle translation rate $a$, respectively. When $a$ and $\delta$ become comparable, $P(m) $ are no longer exponential. 
Rather, they  are peaked at small cluster sizes, indicating that small clusters are more stable. This is because particles cannot easily escape from clusters due to the strong attractive interactions. However, the asymptotic decay remains exponential, as visualized in the inset. This suggests that, although the behavior of small clusters deviates from the exponential form, larger clusters still follow an exponential decay. Fig. \ref{fig:cs}(e) shows the deviation from exponential form.

Contour plots are indeed a powerful tool for visualizing and understanding the clustering behavior of particles, especially when examining the effects of various parameters together. 
They contribute to identify regions where particle clustering behavior changes significantly, indicating non-monotonous behavior. In particular, Fig. ~\ref{fig:cs} (c) displays the overall behaviour of
$\langle m \rangle$ as a function of $J_1$ and $a$ in $\delta^{'} \gg 1$ region. This contour plot shows a non-monotonous re-entrant like behaviour along both axis. The reentrance along $J_1$ is expected due to the competition between  $J_1$ and $J_2$ whereas along $a$  the reentrance emerges due to competition between  $a$ and $\delta$. Fig. ~\ref{fig:cs} (f) displays the $\langle m \rangle$ as a function of $J_1$ and $\delta$ in $Q \gg 1$ region. This contour plot shows that re-entrant like behaviour develops only along $J_1$ axis.  

\subsection{Experimental feasibility and parameter estimation}

{\it In-vitro} experiments performed with microfabricated quasi-1D  substrate have been used to study and quantify the cell-cell interactions \cite{ananyo,cil1,cil2}. In particular, such 
1D substrate with ring geometry has been use for investigating  Madin-Darby Canine Kidney (MDCK) cell aggregates and quantifying the cell-cell interaction strengths \cite{ananyo}. $\rho$ in these studies range from $(0.1 -1.0)$. By comparison of the polarity alignment term of the interaction potential in Ref.\cite{ananyo}, with the alignment term in Eq. (\ref{eqn-H1}), we can estimate the typical values of CIL parameter strengths, $J_1$ and $J_2$. For the sake of simplicity, we assume $J_3 = 0$. In that case $J_1 \sim 1.36$ while $J_2\sim 0.44$. The typical size (spatial extent of these cells along the substrate), $\epsilon \sim 50 ~\mu ms$ and the speed of single cells, $v_o \sim 2 \mu m ~min^{-1}$ \cite{ananyo}. From the stand point of our discrete active spin model, this corresponds to a bare hopping rate, $a \sim 0.04 ~min^{-1}$, where we taken the lattice spacing $\epsilon$ to be identical to the size of the individual cell in the 1D array. It was also observed in the experiments that the rate of polarity switching rate for cell doublets (cluster of two cells) from the configuration to $(\rightarrow ~ \leftarrow)$ to $(\rightarrow ~ \rightarrow)$, $k_p \simeq 0.05$ \cite{ananyo}. We can then estimate the value of bare polarity switching rate, $b = k_p e^{-J_1} \sim 0.01~ min^{-1}$, for our model. Accordingly, the P\'eclet number $Q \sim 4$. While we are not aware of specific estimates of the cell-cell attractive parameter strength, $J_o$, in principle, it can be estimated from the cluster characteristics, e.g., $P(m)$ and $\langle m \rangle$, by the controlled {\it in-vitro} experiments. 

Although we have estimated the parameter range for a particular system (MDCK wild type cells in 1D array)\cite{ananyo}, individual cell-cell interaction parameters can span a range of at least one order of magnitude, depending on cell type and the underlying physiological conditions of specific experiments. Therefore it follows that the collective spatio-temporal characteristics of these cell collectives would display diverse kind of behaviour depending on the specific choice of parameters. Our investigations have spanned this wide spectrum of biologically relevant parameter range of generic cell-cell interactions.  

\section{Discussion}
Using a discrete active spin model, we have  analyzed the impact that attractive and alignment interactions have on the emergent behaviour of cells that self-propel and interact through contact inhibition of locomotion (CIL), in a confined,  quasi-1D geometry. 
While in the absence of vacancies the model can be solved exactly, in the presence of vacancies, the interplay of self-propulsion, attraction and CIL results in richer scenarios that characterize the steady state properties of the system. We have shown that in the absence of  CIL and  alignment interactions,  the model can be exactly mapped to Katz-Lebowitz-Spohn (KLS) model for $Q \ll 1$. We have correspondingly derived an analytic expression for the average attractive energy, average cluster size and cluster size distribution. 
These expressions have allowed us to analyze the deviations in the behavior of the system as we move away from the regime of low $Q$.

Our investigations have revealed two important features of cell assemblies: (a) The interplay of self-propulsion and attraction gives rise to a non-monotonic dependence of cluster sizes. (b) Somewhat counterintuitively, cluster size exhibits a non-monotonic dependence on CIL strength $J_1$, for fixed $J_2$. This highlights the role of the asymmetry of CIL in shaping cellular organization under confinement. 

For  $Q \gg 1$, an approximate expression for cluster size distribution and average cluster size can be obtained by minimizing an effective Helmholtz free energy which includes the energy contributions due to the interactions and configurational entropy \cite{sm-scirep}. This approach, leads to a prediction of an exponential form of the cluster size distribution, a $Q^{\frac{1}{2}}$ dependence of the average cluster size, and a subdominant dependence of the average cluster size on the attractive strength parameter $J_o$. 

The analysis of the topology of the contour plots for the  cluster size reveals that degree of homogenization (corresponding to minimization of average cluster size) in these cell assemblies can be optimized for specific choice of CIL strength and self-propulsion magnitude. 

It is worthwhile pointing out that while coarse-grained hydrodynamic theories of active particle systems with generic self-propulsion and switching have posited and observed motility induced phase separation (MIPS) in 2D systems \cite{active-prl,active-rev}, we have not observed MIPS for the case of our 1D system. Indeed, a systematic analysis of MC simulations for different system sizes revealed that the average cluster size, $\langle m \rangle$, always approached a finite value and in particular it did not scale linearly with system size. This observation is also consistent with the findings of other 1D  active particle systems, e.g., run-and-tumble particles (RTP) model and active Brownian particles (ABP) model in 1D \cite{soto,chandan,goutam}. It is however important to point out two crucial differences of our model with the coarse grained models in Ref.\cite{active-prl,active-rev}: (a) In our case, the particles strictly follow exclusion process, so that they never cross each other, and (b) the scaling regime for the two models are distinct, because we do not scale the switching and hopping parameters either with respect to each other or with respect to system size, unlike in Ref.\cite{active-prl}. Only a careful study of these different scaling regimes (corresponding to different hydrodynamic theories) along with the relaxation of the rule of particle crossing and effect of CIL interaction between clusters can delineate the different conditions under which MIPS can be observed for such 1D active systems. 

The results and insight gained using this model and the identification of the different relevant regimes that we have investigated, provide a solid reference framework to generalize the discrete lattice gas to study the effects of CIL in the context of cellular organization on a 2D substrate as well. 

\begin{acknowledgments}
H.P acknowledges financial support from Mahajyoti fellowship program (MJRF-2022). S.M acknowledges financial support from ANRF-Pair program. S.M also acknowledges the hospitality and financial support for visit to ICTP, Trieste, under the Associateship program. I.P acknowledges support from Ministerio de Ciencia, Innovaci\'on y Universidades MCIU/AEI/FEDER for financial support under grant agreement PID2021-126570NB-100 AEI/FEDER-EU, and Generalitat de Catalunya for financial support under Program Icrea Acad\`emia and project 2021SGR-673.
\end{acknowledgments}

\section*{Author contributions}
H.P carried out simulations, analyzed the  model and data and wrote the manuscript, I.P analyzed model and data and wrote the manuscript. S.M designed study, formulated the theoretical calculations, analyzed the model and data, and wrote the manuscript. All authors reviewed the manuscript. 


\appendix
\section{Simulation Details}
\label{app:simulation-Details}
We perform Monte Carlo (MC) simulations for a system of $N$ particles on $L$ lattice sites, corresponding to an fixed number density $(\rho = N/L)$. The initial configuration is choosing a lattice site at random, with equal probability and placing the N particles. The corresponding polarity of each particle is assigned randomly with equal probability. Subsequently, we use random sequential update procedure wherein we  choose a site at random with equal probability. MC moves for all the processes, e.g., particle hopping and polarity switching is performed with their respective relative rates. We define 1 MC time step as $\frac{N}{k}$, where $k$ is the lowest rate among hopping and polarity switching rate. We wait for an initial duration of 1000 MC time step before we start to collect data for position and polarization of particles so as to ensure that the steady state of the system is attained. Subsequently, we collect the data with a gap of 5 MC time steps to ensure that the samples are uncorrelated. Finally we perform time averaging of the data typically over at $2000-5000$ samples. 
The switching and hopping processes corresponding to different microscopic configurations, and implemented in MC simulations are summarized in Table \ref{table-rates}.

\begin{table*}[t!]
\begin{tabularx}{\textwidth} { 
  | >{\centering\arraybackslash}X 
  | >{\centering\arraybackslash}X 
  | >{\centering\arraybackslash}X | }
 \hline
 The switching dynamics of the particle polarity in the bulk of a cluster & The switching dynamics of particle polarity at the cluster boundary & The translation dynamics of particles with a given polarity  \\
 \hline
 \begin{eqnarray}
\mathrm{\rightarrow~\rightarrow~\rightarrow } ~&\xrightleftharpoons[b e^{ \Delta J/2}]{b e^{ - \Delta J /2}}&~\mathrm{\rightarrow~\leftarrow ~ \rightarrow }\nonumber\\
\mathrm{\leftarrow~\leftarrow~\leftarrow } ~&\xrightleftharpoons[b e^{\Delta J/2}]{b e^{-\Delta J /2}}&~\mathrm{\leftarrow ~ \rightarrow~\leftarrow }\nonumber\\
\mathrm{\rightarrow~\rightarrow~\leftarrow } ~&\xrightleftharpoons[b]{b}&~\mathrm{\rightarrow~\leftarrow ~ \leftarrow }\nonumber\\
\mathrm{\leftarrow~\leftarrow~\rightarrow } ~&\xrightleftharpoons[b]{b}&~\mathrm{\leftarrow~\rightarrow ~ \rightarrow } \nonumber
\end{eqnarray}  & \begin{eqnarray}
\mathrm{\rightarrow~  \rightarrow } ~&\xrightleftharpoons[b  e^{-(J_{2}-2J_{3})/2}]{b e^{ (J_{2}-2J_{3})/2}}&~\mathrm{\leftarrow ~ \rightarrow }\nonumber\\
\mathrm{\rightarrow~  \rightarrow } ~&\xrightleftharpoons[b e^{ (J_{1}+2J_{3})/2}]{b e^{-(J_{1}+2J_{3})/2}}&~\mathrm{\rightarrow ~ \leftarrow }\nonumber\\
\mathrm{\leftarrow~  \leftarrow } ~&\xrightleftharpoons[b e^{-(J_{2}-2J_{3})/2}]{b e^{ (J_{2}-2J_{3})/2}}&~\mathrm{\leftarrow ~ \rightarrow }\nonumber\\
\mathrm{\leftarrow~  \leftarrow } ~&\xrightleftharpoons[be^{(J_{1}+2J_{3})/2}]{b e^{-(J_{1}+2J_{3})/2}}&~\mathrm{\rightarrow ~ \leftarrow } \nonumber
\end{eqnarray}  & \begin{eqnarray}
\mathrm{0~\rightarrow~0~0}~&\xrightarrow{a}&~\mathrm{0~0~\rightarrow~0}\nonumber\\
\mathrm{0~0~\leftarrow~0}~&\xrightarrow{a}&~\mathrm{0~\leftarrow~0~0}\nonumber\\
\mathrm{0~\rightarrow~0~*}~&\xrightarrow{a+\delta}&~\mathrm{0~0~\rightarrow~*}\nonumber\\
\mathrm{*~0~\leftarrow~0}~&\xrightarrow{a+\delta}&~\mathrm{*~\leftarrow~0~0}\nonumber\\
\mathrm{*~\rightarrow~0~0}~&\xrightarrow{a-\delta}&~\mathrm{*~0~\rightarrow~0}\nonumber\\
\mathrm{0~0~\leftarrow~*}~&\xrightarrow{a-\delta}&~\mathrm{0~\leftarrow~0~*}\nonumber\\
\mathrm{*~\rightarrow~0~*}~&\xrightarrow{a}&~\mathrm{*~0~\rightarrow~*}\nonumber\\
\mathrm{*~0~\leftarrow~*}~&\xrightarrow{a}&~\mathrm{*~\leftarrow~0~*}\nonumber
\end{eqnarray}  \\
\hline
\end{tabularx}
\caption{Summary of polarity switching and hopping processes on the lattice: In the absence of any interactions, $a$ correspond to the hopping rate of particle while $b$ corresponds to polarity switching rate of particles in the absence of any interactions. The rate $\delta$ is a measure of attractive interaction strength. Here $\Delta J = J_1 - J_2 + 4 J_3$. $0$ corresponds to an empty lattice site and $*$ corresponds to a particle of either polarity ( Left or Right).}
\label{table-rates}
\end{table*}

\section{Average energy of a finite size cluster} 
\label{app:calculations-ae}

The expression for average energy per particle due to polarity interaction in the thermodynamics limit of $N\rightarrow \infty $ is given by Eq. (\ref{eqn-avgE-CIL}). This expression was obtained by using the form of the partition function $Z$ in the limit of $N\rightarrow \infty$. In general, the partition function is given by $Z_N = \lambda_{1}^{N} + \lambda_{2}^{N}$,  where $\lambda_{1}$ and $\lambda_{2}$ are the large and small eigenvalues of the transfer matrix, respectively. For small clusters, the contribution from the smaller of the two  eigenvalues, i.e.,  $\lambda_{2}$ is no longer negligible and therefore we need to use the expression for the full expression of the partition function for calculating the energy of a finite size cluster of size $m$.  In order to simplify the analysis, we incorporate the effect of CIL through $J_1$ alone by setting $J_2 = 0$. We also set $J_3 = 0$.  The general form of the average energy of an $m$ size cluster in the presence of interparticle attractive interaction has a form,
\begin{equation}
 E(m) = E_{p}(m)+E_{a}(m) \nonumber
\end{equation}

\begin{equation}
 E(m) = m \frac{J_{1}}{2}\frac{ 1}{ \left ( 1 + e^{\frac{J_{1}}{2}} \right )}\left[1- F(m) \right] -(m-1)J_o
\end{equation}
Here, the first term corresponds to contribution due to CIL and alignment interaction between particle polarities while the second term is contribution of attractive interaction. Here, the explicit expression of $F(m)$ is, 
\begin{equation}
F(m) = \frac{ 2 e^{ \frac{J_1}{2}}\left ( e^{\frac{J_1}{2}} -1 \right )^{m-1} }{\left (e^{ \frac{J_1}{2}} +1\right)^{m}+\left (e^{ \frac{J_1}{2}} -1\right)^{m}}
\end{equation}
Note that $F(m) \rightarrow  0$ in the thermodynamic limit of $m \rightarrow  \infty$.

Although here we provided the expression for a special case, the procedure of using Transfer Matrix technique can as well be used to obtain an explicit form of average energy even when $J_2 \neq 0$ and $J_3 \neq 0$. 

\section{Calculation of cluster size by minimization of $\mathcal{F}$}
\label{app:calculations-acs}

\begin{figure*}[t]
    \centering
    \includegraphics[width = \linewidth]{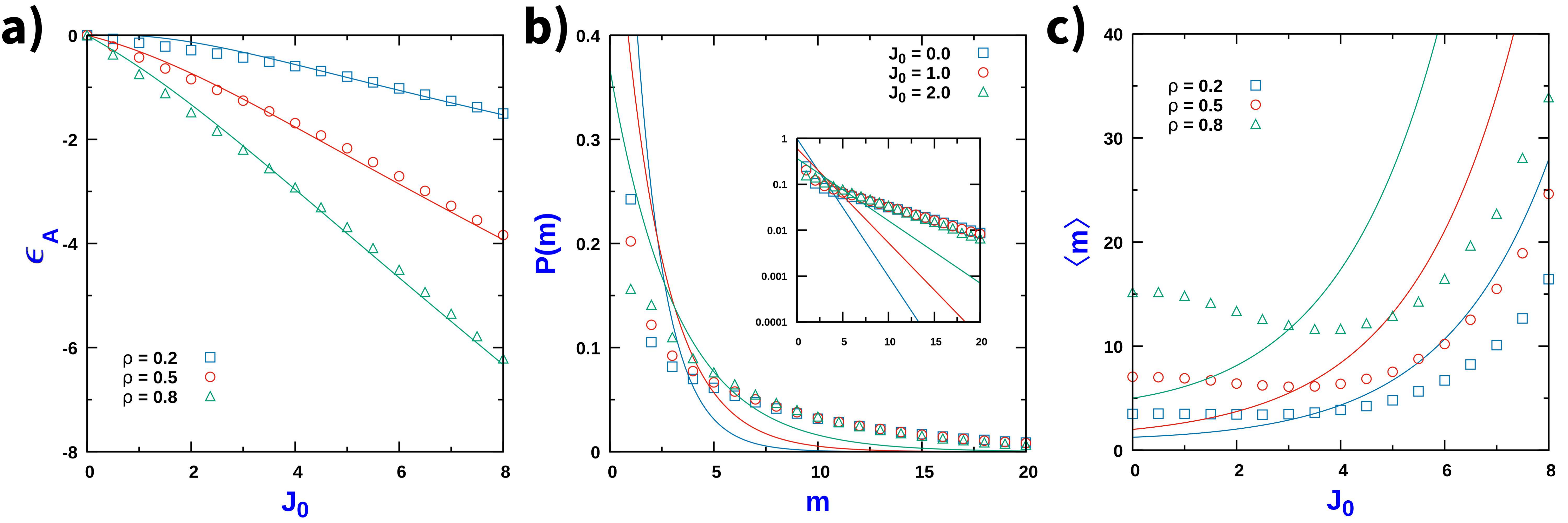}
    \caption{\textbf{(a)} Variation of average attractive energy per site $\epsilon_A=\frac{\left \langle E \right \rangle_A}{L}$ as function of $J_o = \ln\left(\frac{a+\delta}{a-\delta}\right)$ in $Q \gg 1$ $(Q = 50)$: i) $\rho = 0.2$, ii) $\rho = 0.5$, iii) $\rho = 0.8$. \textbf{(b)} Corresponding Cluster size distributions $P(m)$ for different $J_o$ for $\rho = 0.5$. Inset figure is corresponding log plot. \textbf{(c)} Corresponding average cluster size $\left \langle m \right \rangle$ as function of $J_o$. The solid line in (a) corresponds to the expression in Eq. (\ref{eqn-avgE-KLS}) while points correspond to MC simulations. The solid line in (b) corresponds to the expression in Eq. (\ref{eqn-CSD-KLS}) while points correspond to MC simulations. The solid line in (c) corresponds to the expression in Eq. (\ref{eqn-avgCS-KLS}) while points correspond to MC simulations. For all cases $J_1 = J_2 = J_3 = 0$, $a=50$, $b=1$ and MC simulation was done with $L=2000$ and averaging was done over $2000$ samples.}
    \label{fig:KLS-highQ}
\end{figure*}

\begin{figure*}[t]
 \centering
    \includegraphics[width= \linewidth, height = 10cm]{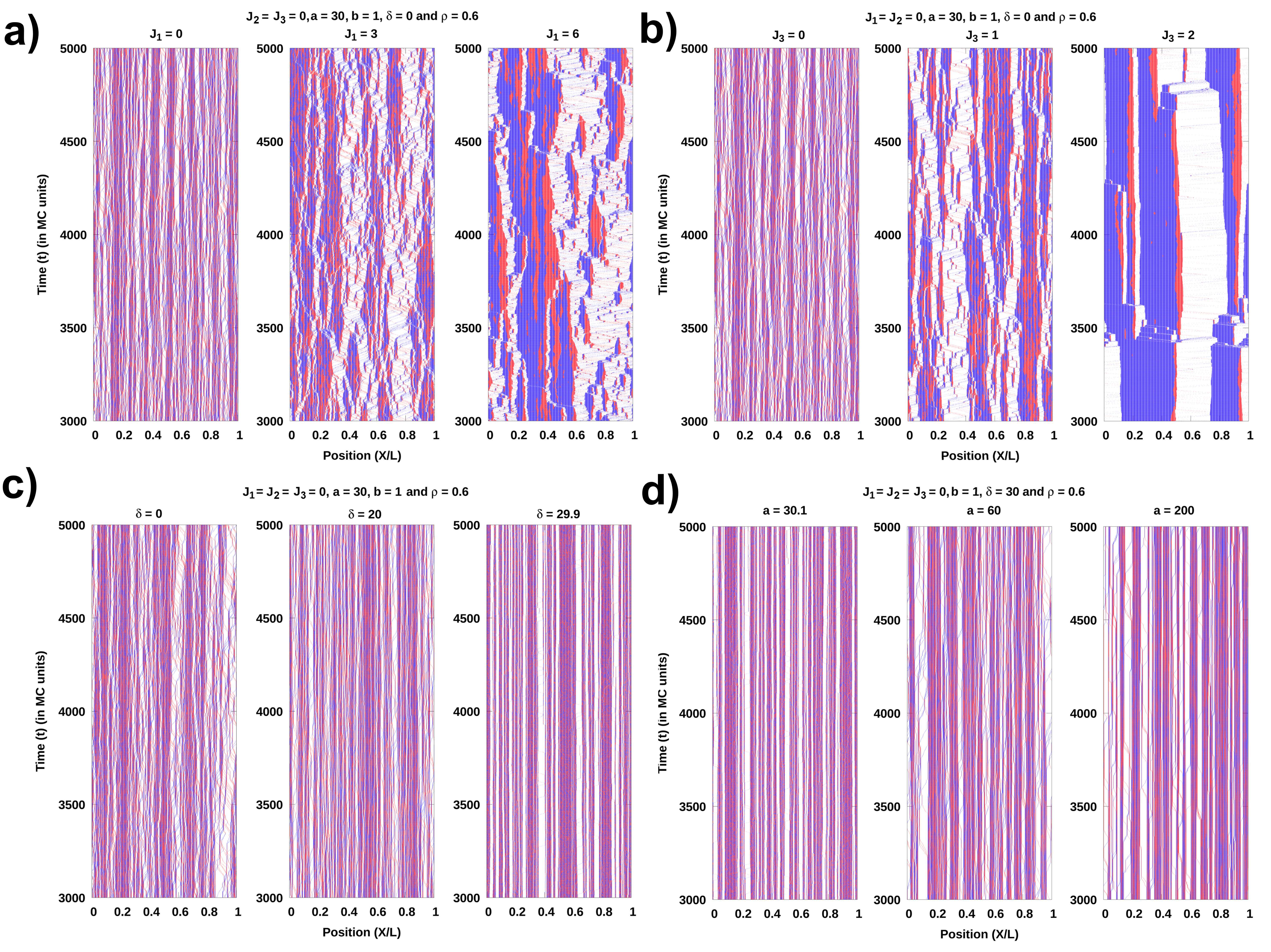}
    \caption{\textbf{Spatio-temporal plot:} Time snapshots of distribution of right polarized $(\rightarrow)$(blue) and left polarized $(\leftarrow)$(red) particles on the lattice. \textbf{(a)} For different $J_1$ in $Q \gg 1$ and no attraction regime $(a = 30, \delta = 0)$: i) $J_1 = 0$, ii) $J_1 = 3$, iii) $J_1 = 6$. Here, $J_2 = 0$ and $J_3 = 0$. \textbf{(b)} For different $J_3$ in $Q \gg 1$ and no attraction regime $(a = 30, \delta = 0)$: i) $J_3 = 0$, ii) $J_3 = 1$, iii) $J_3 = 2$. Here, $J_1 = 0$ and $J_2 = 0$. \textbf{(c)} For different $\delta$ in $Q \gg 1$ and no CIL  and alignment regime $(a = 30, J_1 = 0, J_2 =, J_3 =0)$: i) $\delta = 0$, ii) $\delta = 20$, iii) $\delta = 29.9$. \textbf{(d)} For different $a$ in $\delta^{'} \gg 1$ and no CIL and alignment regime $(\delta = 30, J_1 = 0, J_2 =, J_3 =0)$: i) $a = 30.1$, ii) $a = 60$, iii) $a = 200$. For all cases particle number density $\rho =0.6$, $b=1$ and MC simulation was done with $L=1000$.} 
    \label{fig:morphology}
\end{figure*}

In this appendix section, we provide the steps of derivation of the average cluster size and cluster size distribution function in the dense cluster phase using the procedure of minimizing the Effective Helmholtz Free energy $ \mathcal F$ in the limit of $ Q\gg 1$. 
Using the expression of $ \mathcal F$ from Eq. (\ref{eqn-F}), the expression for $\delta \mathcal{F}$ assumes a form,
\begin{eqnarray}
\delta \mathcal{F}=\sum_{m} m\left \{ \frac{J_{1}}{2}\frac{ 1}{ \left ( 1 + e^{\frac{ J_{1}}{2}} \right )}\left[ 1- F(m) \right]\right \} \delta G_{c}(m) \nonumber\\
+\sum_{m}  \left[\ln G_{c}(m) -m(J_o +\lambda)-(\lambda-J_o)\right] \delta G_{c}(m) 
\end{eqnarray}

By setting $\delta \mathcal{F} = 0$ we can obtain, form of $G_c(m)$ no of clusters of cluster size $m$ in dense cluster phase,
\begin{equation}
G_{c}(m)=A_{c}e^{-\frac{m}{m_{c}}}e^{\frac{\mathcal{F}(m)m}{m_{c_{1}}}} 
\label{eqn-Gc1}
\end{equation}
Where,
\begin{align}
A_{c}&=e^{\left \{ \gamma-J_o \right \}} \nonumber \\
 m_{c}&= \frac{1}{\left \{ \frac{J_{1}}{2}\frac{ 1}{ \left ( 1 + e^{\frac{ J_{1}}{2}} \right )} - J_o -\lambda  \right \}} \nonumber \\
 m_{c_{1}}&= \frac{1}{\left \{ \frac{J_{1}}{2}\frac{ 1}{ \left ( 1 + e^{\frac{ J_{1}}{2}} \right )} \right \}} \nonumber   
\end{align}
In the thermodynamic limit of $m \rightarrow  \infty$, note that $F(m) \rightarrow  0$ therefore we have expression of $G_c(m)$ as,

\begin{equation}
G_{c}(m)=A_{c}e^{-\frac{m}{m_{c}}} 
\label{eqn-Gc2}
\end{equation}

In the limit of no attractive interactions $J_o = 0$, the expression of $G_c(m)$ reduces to the form presented in the Ref. \cite{sm-scirep}.

Similar argument for the gas phase yields exponential form of $G_{g}(m)$ i.e. no of $m$ size gas phases,
\begin{equation}
G_{g}(m)=A_{g}e^{-\frac{m}{m_{g}}} 
\label{eqn-Gg}
\end{equation}
In order to fix the parameters in Eq. (\ref{eqn-Gc2}) and (\ref{eqn-Gg}), $A_{c}$, $A_{g}$, $m_{c}$ and $m_{g}$ that characterize univocally the coexisting cluster size distributions of the gas and dense cluster phase using appropriate conditions taken into accounts.\\
a) The total number of gas phases must equal to the total number of clusters in the dense cluster phase.
\begin{equation}
\sum_{m} G_{c}(m) = \sum_{m} G_{g}(m) = \Omega\nonumber
\label{condition1}\tag{a}
\end{equation}
b) The total number of sites occupied by the gas phases and the total number sites occupied by clusters in the dense cluster phases must equal to the number of lattice sites.
\begin{equation}
\sum_{m} m G_{c}(m) + \sum_{m} m G_{g}(m) = L\nonumber
\label{condition2}\tag{b}
\end{equation}
\noindent
c) If $\phi_{c}$ and $\phi_{g}$ are the number density of particles in dense cluster region and gas region, with $\left \langle m_{c} \right \rangle$ and $\left \langle m_{g} \right \rangle$ being the average length of the dense cluster and gas region, then the overall particle density $\rho$ should obey,
\begin{equation}
\left \langle m_{c}\right \rangle \phi_{c} + \left \langle m_{g}\right \rangle \phi_{g}=\left ( \left \langle m_{c}\right \rangle + \left \langle m_{g}\right \rangle \right ) \rho \nonumber 
\label{condition3}\tag{c}
\end{equation}
\noindent
d) If the hopping rate is much larger than the switching rate, the gas region has typically a very low density of particles, hence $\phi_{g} \ll 1$. When a switching event at the boundary of a dense cluster occurs, a particle is emitted into the gas region. This leads to the production of a dimer within the gas region which are the dominant clusters in the gas phase. Accordingly, we invoke the condition that in the limit of $Q \gg 1$, the steady state is determined by the condition of matching the dimer production and disintegration rates.

The dimer production in the gas occurs when a particle at the edge of the adjoining dense cluster (with its polarity pointing inwards towards the bulk of cluster), has flipped its polarity and thus breaks away into the gas region, and within the time $\tau_0$ that it takes to reach the adjacent dense cluster.
The average time that it takes for a particle on the edge of the dense cluster to reach the edge of the adjacent dense cluster reads $\left \langle \tau_0 \right \rangle = \frac{\left \langle m_{g} \right \rangle}{a}$. Assuming that the neighbors of the edge particle of the dense cluster are pointing inwards, the rate at which the particle comes out of the cluster is the geometric mean of the rate of switching polarity and the rate of hopping out of the cluster. Since switching rate is $b$ and hopping out rate $a- \delta$.
\begin{equation}
k_{out} = \frac{b(a-\delta)}{b+(a-\delta)}\nonumber
\end{equation}
\noindent
The total number of gas phases are $\sum_{m} G_{g}(m)$. The overall production rate of dimmer is approximately,
\begin{equation}
W_{p}^{2}=2 \left ( \frac{b(a-\delta)}{b+(a-\delta)}\right )^{2}\frac{\left \langle m_{g} \right \rangle}{a} \sum_{m} G_{g}(m) \nonumber  
\end{equation}
The disintegration of a dimer would occur when either of the particles of the dimer cluster switches their direction of polarity and comes out of the cluster. The rate at which the particle comes out of the cluster is the geometric mean of the rate of switching polarity and the rate of hopping out of the cluster. While in the absence of CIL the switching rate is $b$, in the presence of CIL the switching rate is $be^{\frac{J_1}{2}}$ and hopping out rate $a- \delta$.
\begin{equation}
k^{'}_{out} = \frac{be^{\frac{J_{1}}{2}}(a-\delta)}{be^{\frac{J_{1}}{2}}+(a-\delta)}\nonumber
\end{equation}
\noindent
Since the number of such dimer clusters is simply $P_{c}(2)$. Then overall disintegration rate of dimmer is approximately, 
\begin{equation}
W_{d}^{2}=2 \left ( \frac{be^{\frac{J_{1}}{2}}(a-\delta)}{be^{\frac{J_{1}}{2}}+(a-\delta)} \right ) G_{c}(2) \nonumber
\end{equation}
After balancing the dimer production and disintegration rates, we arrive at the following condition,
\begin{equation}
\left ( \frac{b(a-\delta)}{b+(a-\delta)}\right )^{2}\frac{\left \langle m_{g} \right \rangle}{a} \sum_{m} G_{g}(m)= \left ( \frac{be^{\frac{J_{1}}{2}}(a-\delta)}{be^{\frac{J_{1}}{2}}+(a-\delta)} \right ) G_{c}(2) 
\label{condition4}\tag{d}
\end{equation}
Since the particle flux from the gas into the cluster region, $\frac{(a+\delta)\phi_{g}}{2}$, must equal the particle flux from the dense cluster region, $\frac{b(a-\delta)}{b+(a-\delta)}$ (due to particle switching at the cluster boundary and breaks away). At steady state, we arrive at the condition, 
\begin{equation}
\phi_{g}=\frac{2\left ( \frac{b(a-\delta)}{b+(a-\delta)} \right )}{a+\delta}
\label{condition5}\tag{e}
\end{equation}

We approximate the sums in Eq. (\ref{condition1}), Eq. (\ref{condition2}) and Eq. (\ref{condition4}) by integrals to obtain an approximate expression for the mean cluster size. We set the integration limit for the dense cluster phase between $2$ and $\infty$ since, for having a dense cluster, there needs to be at least $2$ particles. For the gas phase, there has to be at least $1$ site, which sets the lower integration limit. The upper integration limit is taken to $\infty$. With these conversions we have,
\begin{equation}
 \sum_{m} G_{c}(m) = \int_{2}^{\infty} G_{c}(m) dm = A_{c}  m_{c}  e^{-\frac{2}{m_{c}}} \nonumber   
\end{equation}
\begin{equation}
\sum_{m} G_{g}(m) = \int_{1}^{\infty} G_{g}(m) dm = A_{g}  m_{g}  e^{-\frac{1}{m_{g}}} \nonumber
\end{equation}
\begin{equation}
\sum_{m} m G_{c}(m) = \int_{2}^{\infty} m G_{c}(m) dm = A_{c}  m_{c}  e^{-\frac{2}{m_{c}}} \left ( 2 + m_{c} \right ) \nonumber   
\end{equation}
\begin{equation}
\sum_{m} m G_{g}(m) = \int_{1}^{\infty} m G_{g}(m) dm = A_{g}  m_{g}  e^{-\frac{1}{m_{g}}} \left ( 1 + m_{g} \right ) \nonumber
\end{equation}
This approximations leads to, 
\begin{eqnarray}
\left \langle m_c \right \rangle&=&2+m_{c} \nonumber \\
\left \langle m_g \right \rangle&=&1+m_g \nonumber
\end{eqnarray}
Substituting these integral form in Eq. (\ref{condition1}), Eq. (\ref{condition2}), Eq. (\ref{condition3}), and Eq. (\ref{condition4}) we get relations involving constants $A_{c}$, $A_{g}$, $m_{c}$ and $m_{g}$. Finally we arrive at the expression for average cluster size in dense cluster phase,
\begin{widetext}
\begin{equation}
\left \langle m_c \right \rangle=1+\sqrt{1+\left \{ \frac{\rho ab}{(a-\delta)}+2\rho a+\frac{\rho a(a-\delta)}{b}-\frac{2ab}{(a+\delta)}-\frac{2a(a-\delta)}{(a+\delta)} \right \}\frac{e^{\frac{J_{1}}{2}}}{(1-\rho)\left ( be^{\frac{J_{1}}{2}}+(a-\delta) \right )}} \nonumber
\end{equation}
In terms of dimensionless quantities $Q=\frac{a}{b}$ and $J_o=\ln\left(\frac{a+\delta}{a-\delta}\right)$ we can rewrite the expression of avarage size of cluster as,

\begin{equation}
\left\langle m_c \right\rangle = 1 + \sqrt{1 + \left[\frac{\left( \frac{\rho}{1 - T} - \frac{2}{1 + T} \right)}{Q} + \left( 2\rho - 2e^{-J_o} \right) + \rho (1 - T)Q\right]\frac{e^{\frac{J_1}{2}}}{\left( 1 - \rho \right)\left( \dfrac{e^{\frac{J_1}{2}}}{Q} + (1 - T) \right)}} \nonumber
\end{equation}
where, $T=\tanh\left ( \frac{J_o}{2} \right )$. 
\end{widetext}

\section{Additional figures} 
\label{app:additional-figures}

\subsection{Comparison of KLS model cluster characteristics with MC simulations in $Q \gg 1$ limit}

Fig. \ref{fig:KLS-highQ} shows the comparison of cluster characteristics using MC simulation results in the $Q \gg 1$ limit, with analytical form of average energy per site, P(m) and $\langle m \rangle$ obtained by mapping the problem to KLS model. As expected, in the $Q\gg 1$ limit, the divergence of the MC simulation results from the KLS model results are drastic. However we find the analytical expression of the average energy per cluster still matches very well the MC simulation results.

\subsection{Spatio-Temporal Plots}

In this appendix, we present additional spatio-temporal plots (see Fig. {\ref{fig:morphology}}) to further illustrate the dynamic behavior of the system under investigation. These plots provide a more comprehensive view of the evolution of spatial patterns over time. By analyzing these visualizations, we aim to highlight features that may not be apparent from time-averaged or purely spatial representations, thereby offering deeper insights into the underlying processes.

\bibliography{paper-CIL-PR-E-v3}

\providecommand{\noopsort}[1]{}\providecommand{\singleletter}[1]{#1}%
\begin{thebibliography}{36}%
\makeatletter
\providecommand \@ifxundefined [1]{%
 \@ifx{#1\undefined}
}%
\providecommand \@ifnum [1]{%
 \ifnum #1\expandafter \@firstoftwo
 \else \expandafter \@secondoftwo
 \fi
}%
\providecommand \@ifx [1]{%
 \ifx #1\expandafter \@firstoftwo
 \else \expandafter \@secondoftwo
 \fi
}%
\providecommand \natexlab [1]{#1}%
\providecommand \enquote  [1]{``#1''}%
\providecommand \bibnamefont  [1]{#1}%
\providecommand \bibfnamefont [1]{#1}%
\providecommand \citenamefont [1]{#1}%
\providecommand \href@noop [0]{\@secondoftwo}%
\providecommand \href [0]{\begingroup \@sanitize@url \@href}%
\providecommand \@href[1]{\@@startlink{#1}\@@href}%
\providecommand \@@href[1]{\endgroup#1\@@endlink}%
\providecommand \@sanitize@url [0]{\catcode `\\12\catcode `\$12\catcode
  `\&12\catcode `\#12\catcode `\^12\catcode `\_12\catcode `\%12\relax}%
\providecommand \@@startlink[1]{}%
\providecommand \@@endlink[0]{}%
\providecommand \url  [0]{\begingroup\@sanitize@url \@url }%
\providecommand \@url [1]{\endgroup\@href {#1}{\urlprefix }}%
\providecommand \urlprefix  [0]{URL }%
\providecommand \Eprint [0]{\href }%
\providecommand \doibase [0]{http://dx.doi.org/}%
\providecommand \selectlanguage [0]{\@gobble}%
\providecommand \bibinfo  [0]{\@secondoftwo}%
\providecommand \bibfield  [0]{\@secondoftwo}%
\providecommand \translation [1]{[#1]}%
\providecommand \BibitemOpen [0]{}%
\providecommand \bibitemStop [0]{}%
\providecommand \bibitemNoStop [0]{.\EOS\space}%
\providecommand \EOS [0]{\spacefactor3000\relax}%
\providecommand \BibitemShut  [1]{\csname bibitem#1\endcsname}%
\let\auto@bib@innerbib\@empty
\bibitem [{\citenamefont {Smeets}\ and\ \citenamefont
  {et~al.}(2016)}]{igna-pnas}%
  \BibitemOpen
  \bibfield  {author} {\bibinfo {author} {\bibfnamefont {B.}~\bibnamefont
  {Smeets}}\ and\ \bibinfo {author} {\bibnamefont {et~al.}},\ }\href@noop {}
  {\bibfield  {journal} {\bibinfo  {journal} {Proc. Natl. Acad. Sci.}\ }\textbf
  {\bibinfo {volume} {113}},\ \bibinfo {pages} {14621} (\bibinfo {year}
  {2016})}\BibitemShut {NoStop}%
\bibitem [{\citenamefont {Alert}\ and\ \citenamefont {Trepat}(2020)}]{trepat}%
  \BibitemOpen
  \bibfield  {author} {\bibinfo {author} {\bibfnamefont {R.}~\bibnamefont
  {Alert}}\ and\ \bibinfo {author} {\bibfnamefont {X.}~\bibnamefont {Trepat}},\
  }\href@noop {} {\bibfield  {journal} {\bibinfo  {journal} {Annu. rev. Conens.
  Matter Phys.}\ }\textbf {\bibinfo {volume} {11}},\ \bibinfo {pages} {77}
  (\bibinfo {year} {2020})}\BibitemShut {NoStop}%
\bibitem [{\citenamefont {Friedl}\ and\ \citenamefont
  {Gilmour}(2009)}]{ignaref2}%
  \BibitemOpen
  \bibfield  {author} {\bibinfo {author} {\bibfnamefont {P.}~\bibnamefont
  {Friedl}}\ and\ \bibinfo {author} {\bibfnamefont {D.}~\bibnamefont
  {Gilmour}},\ }\href@noop {} {\bibfield  {journal} {\bibinfo  {journal} {Nat.
  Rev. Mol. Cell. Biol.}\ }\textbf {\bibinfo {volume} {10}},\ \bibinfo {pages}
  {445} (\bibinfo {year} {2009})}\BibitemShut {NoStop}%
\bibitem [{\citenamefont {Moh}\ and\ \citenamefont {Shen}(2009)}]{cancer}%
  \BibitemOpen
  \bibfield  {author} {\bibinfo {author} {\bibfnamefont {M.~C.}\ \bibnamefont
  {Moh}}\ and\ \bibinfo {author} {\bibfnamefont {S.}~\bibnamefont {Shen}},\
  }\href@noop {} {\bibfield  {journal} {\bibinfo  {journal} {Cell Adh Migr.}\
  }\textbf {\bibinfo {volume} {3}},\ \bibinfo {pages} {334} (\bibinfo {year}
  {2009})}\BibitemShut {NoStop}%
\bibitem [{\citenamefont {Santos}\ and\ \citenamefont {Liberali}(2019)}]{FEBS}%
  \BibitemOpen
  \bibfield  {author} {\bibinfo {author} {\bibfnamefont {A.~X.~S.}\
  \bibnamefont {Santos}}\ and\ \bibinfo {author} {\bibfnamefont
  {P.}~\bibnamefont {Liberali}},\ }\href@noop {} {\bibfield  {journal}
  {\bibinfo  {journal} {FEBS J.}\ }\textbf {\bibinfo {volume} {286(8)}},\
  \bibinfo {pages} {1495} (\bibinfo {year} {2019})}\BibitemShut {NoStop}%
\bibitem [{\citenamefont {Tambe}\ and\ \citenamefont
  {et~al.}(2011)}]{levineref14}%
  \BibitemOpen
  \bibfield  {author} {\bibinfo {author} {\bibfnamefont {D.~T.}\ \bibnamefont
  {Tambe}}\ and\ \bibinfo {author} {\bibnamefont {et~al.}},\ }\href@noop {}
  {\bibfield  {journal} {\bibinfo  {journal} {Nat. Mater.}\ }\textbf {\bibinfo
  {volume} {10}},\ \bibinfo {pages} {469} (\bibinfo {year} {2011})}\BibitemShut
  {NoStop}%
\bibitem [{\citenamefont {Ramaswamy}(2010)}]{sriram-rev}%
  \BibitemOpen
  \bibfield  {author} {\bibinfo {author} {\bibfnamefont {S.}~\bibnamefont
  {Ramaswamy}},\ }\href@noop {} {\bibfield  {journal} {\bibinfo  {journal}
  {Ann. Rev. Cond. Mat. Phys.}\ }\textbf {\bibinfo {volume} {1}},\ \bibinfo
  {pages} {323} (\bibinfo {year} {2010})}\BibitemShut {NoStop}%
\bibitem [{\citenamefont {Ladoux}\ and\ \citenamefont {Mege}(2017)}]{cadherin}%
  \BibitemOpen
  \bibfield  {author} {\bibinfo {author} {\bibfnamefont {B.}~\bibnamefont
  {Ladoux}}\ and\ \bibinfo {author} {\bibfnamefont {R.}~\bibnamefont {Mege}},\
  }\href@noop {} {\bibfield  {journal} {\bibinfo  {journal} {Nat. Rev. Mol.
  Cell Biol.}\ }\textbf {\bibinfo {volume} {18(12)}},\ \bibinfo {pages} {743}
  (\bibinfo {year} {2017})}\BibitemShut {NoStop}%
\bibitem [{\citenamefont {Roycroft}\ and\ \citenamefont {Mayor}(2015)}]{cil1}%
  \BibitemOpen
  \bibfield  {author} {\bibinfo {author} {\bibfnamefont {A.}~\bibnamefont
  {Roycroft}}\ and\ \bibinfo {author} {\bibfnamefont {R.}~\bibnamefont
  {Mayor}},\ }\href@noop {} {\bibfield  {journal} {\bibinfo  {journal} {Trends
  in Cell Biol.}\ }\textbf {\bibinfo {volume} {25(7)}},\ \bibinfo {pages} {373}
  (\bibinfo {year} {2015})}\BibitemShut {NoStop}%
\bibitem [{\citenamefont {Scarpa}\ and\ \citenamefont {et~al.}(2013)}]{cil2}%
  \BibitemOpen
  \bibfield  {author} {\bibinfo {author} {\bibfnamefont {E.}~\bibnamefont
  {Scarpa}}\ and\ \bibinfo {author} {\bibnamefont {et~al.}},\ }\href@noop {}
  {\bibfield  {journal} {\bibinfo  {journal} {Biol. Open}\ }\textbf {\bibinfo
  {volume} {2(9)}},\ \bibinfo {pages} {901} (\bibinfo {year}
  {2013})}\BibitemShut {NoStop}%
\bibitem [{\citenamefont {D.A.~Kulawiak}\ and\ \citenamefont
  {Rappel}(2016)}]{cil-plos}%
  \BibitemOpen
  \bibfield  {author} {\bibinfo {author} {\bibfnamefont {B.~A.~C.}\
  \bibnamefont {D.A.~Kulawiak}}\ and\ \bibinfo {author} {\bibfnamefont {W.-J.}\
  \bibnamefont {Rappel}},\ }\href@noop {} {\bibfield  {journal} {\bibinfo
  {journal} {PloS Comput Biol}\ }\textbf {\bibinfo {volume} {12}},\ \bibinfo
  {pages} {e1005239} (\bibinfo {year} {2016})}\BibitemShut {NoStop}%
\bibitem [{\citenamefont {Zimmermann}\ \emph {et~al.}(2016)\citenamefont
  {Zimmermann}, \citenamefont {Camley}, \citenamefont {Rappel},\ and\
  \citenamefont {Levine}}]{levine-pnas}%
  \BibitemOpen
  \bibfield  {author} {\bibinfo {author} {\bibfnamefont {J.}~\bibnamefont
  {Zimmermann}}, \bibinfo {author} {\bibfnamefont {B.~A.}\ \bibnamefont
  {Camley}}, \bibinfo {author} {\bibfnamefont {W.}~\bibnamefont {Rappel}}, \
  and\ \bibinfo {author} {\bibfnamefont {H.}~\bibnamefont {Levine}},\
  }\href@noop {} {\bibfield  {journal} {\bibinfo  {journal} {Proc. Natl. Acad.
  Sci.}\ }\textbf {\bibinfo {volume} {113}},\ \bibinfo {pages} {2660} (\bibinfo
  {year} {2016})}\BibitemShut {NoStop}%
\bibitem [{\citenamefont {Cates}\ and\ \citenamefont
  {Tailleur}(2015)}]{active-rev}%
  \BibitemOpen
  \bibfield  {author} {\bibinfo {author} {\bibfnamefont {M.~E.}\ \bibnamefont
  {Cates}}\ and\ \bibinfo {author} {\bibfnamefont {J.}~\bibnamefont
  {Tailleur}},\ }\href@noop {} {\bibfield  {journal} {\bibinfo  {journal} {Ann.
  Rev. Cond. Mat. Phys.}\ }\textbf {\bibinfo {volume} {6}},\ \bibinfo {pages}
  {219} (\bibinfo {year} {2015})}\BibitemShut {NoStop}%
\bibitem [{\citenamefont {Szabo}\ and\ \citenamefont {et~al.}(2006)}]{agent1}%
  \BibitemOpen
  \bibfield  {author} {\bibinfo {author} {\bibfnamefont {B.}~\bibnamefont
  {Szabo}}\ and\ \bibinfo {author} {\bibnamefont {et~al.}},\ }\href@noop {}
  {\bibfield  {journal} {\bibinfo  {journal} {Phys. Rev. E.}\ }\textbf
  {\bibinfo {volume} {74}},\ \bibinfo {pages} {061908} (\bibinfo {year}
  {2006})}\BibitemShut {NoStop}%
\bibitem [{\citenamefont {Belmonte}\ \emph {et~al.}(2008)\citenamefont
  {Belmonte}, \citenamefont {Thomas}, \citenamefont {Brunnet}, \citenamefont
  {de~Almeida},\ and\ \citenamefont {Chate}}]{agent2}%
  \BibitemOpen
  \bibfield  {author} {\bibinfo {author} {\bibfnamefont {J.~M.}\ \bibnamefont
  {Belmonte}}, \bibinfo {author} {\bibfnamefont {G.~L.}\ \bibnamefont
  {Thomas}}, \bibinfo {author} {\bibfnamefont {L.~G.}\ \bibnamefont {Brunnet}},
  \bibinfo {author} {\bibfnamefont {R.~M.~C.}\ \bibnamefont {de~Almeida}}, \
  and\ \bibinfo {author} {\bibfnamefont {H.}~\bibnamefont {Chate}},\
  }\href@noop {} {\bibfield  {journal} {\bibinfo  {journal} {Phys. Rev. Lett.}\
  }\textbf {\bibinfo {volume} {100}},\ \bibinfo {pages} {248702} (\bibinfo
  {year} {2008})}\BibitemShut {NoStop}%
\bibitem [{\citenamefont {Camley}\ \emph {et~al.}(2016)\citenamefont {Camley},
  \citenamefont {Zimmermann}, \citenamefont {Levine},\ and\ \citenamefont
  {Rappel}}]{levine-prl}%
  \BibitemOpen
  \bibfield  {author} {\bibinfo {author} {\bibfnamefont {B.~A.}\ \bibnamefont
  {Camley}}, \bibinfo {author} {\bibfnamefont {J.}~\bibnamefont {Zimmermann}},
  \bibinfo {author} {\bibfnamefont {H.}~\bibnamefont {Levine}}, \ and\ \bibinfo
  {author} {\bibfnamefont {W.}~\bibnamefont {Rappel}},\ }\href@noop {}
  {\bibfield  {journal} {\bibinfo  {journal} {Phys. Rev. Lett.}\ }\textbf
  {\bibinfo {volume} {116}},\ \bibinfo {pages} {098101} (\bibinfo {year}
  {2016})}\BibitemShut {NoStop}%
\bibitem [{\citenamefont {Soto}\ and\ \citenamefont
  {Golestanian}(2014)}]{soto}%
  \BibitemOpen
  \bibfield  {author} {\bibinfo {author} {\bibfnamefont {R.}~\bibnamefont
  {Soto}}\ and\ \bibinfo {author} {\bibfnamefont {R.}~\bibnamefont
  {Golestanian}},\ }\href@noop {} {\bibfield  {journal} {\bibinfo  {journal}
  {Phys. Rev. E}\ }\textbf {\bibinfo {volume} {89}},\ \bibinfo {pages} {012706}
  (\bibinfo {year} {2014})}\BibitemShut {NoStop}%
\bibitem [{\citenamefont {Duclos}\ and\ \citenamefont {et~al.}(2018)}]{hydro1}%
  \BibitemOpen
  \bibfield  {author} {\bibinfo {author} {\bibfnamefont {G.}~\bibnamefont
  {Duclos}}\ and\ \bibinfo {author} {\bibnamefont {et~al.}},\ }\href@noop {}
  {\bibfield  {journal} {\bibinfo  {journal} {Nat. Phys.}\ }\textbf {\bibinfo
  {volume} {14(7)}},\ \bibinfo {pages} {728} (\bibinfo {year}
  {2018})}\BibitemShut {NoStop}%
\bibitem [{\citenamefont {Saw}\ and\ \citenamefont {et~al.}(2017)}]{hydro2}%
  \BibitemOpen
  \bibfield  {author} {\bibinfo {author} {\bibfnamefont {T.~B.}\ \bibnamefont
  {Saw}}\ and\ \bibinfo {author} {\bibnamefont {et~al.}},\ }\href@noop {}
  {\bibfield  {journal} {\bibinfo  {journal} {Nature}\ }\textbf {\bibinfo
  {volume} {544}},\ \bibinfo {pages} {212} (\bibinfo {year}
  {2017})}\BibitemShut {NoStop}%
\bibitem [{\citenamefont {Bertrand}\ and\ \citenamefont
  {et~al.}(2024)}]{ananyo}%
  \BibitemOpen
  \bibfield  {author} {\bibinfo {author} {\bibfnamefont {T.}~\bibnamefont
  {Bertrand}}\ and\ \bibinfo {author} {\bibnamefont {et~al.}},\ }\href@noop {}
  {\bibfield  {journal} {\bibinfo  {journal} {Phys. Rev. Res.}\ }\textbf
  {\bibinfo {volume} {6}},\ \bibinfo {pages} {023022} (\bibinfo {year}
  {2024})}\BibitemShut {NoStop}%
\bibitem [{\citenamefont {Potdar}\ \emph {et~al.}(2023)\citenamefont {Potdar},
  \citenamefont {Pagonabarraga},\ and\ \citenamefont {Muhuri}}]{sm-scirep}%
  \BibitemOpen
  \bibfield  {author} {\bibinfo {author} {\bibfnamefont {H.}~\bibnamefont
  {Potdar}}, \bibinfo {author} {\bibfnamefont {I.}~\bibnamefont
  {Pagonabarraga}}, \ and\ \bibinfo {author} {\bibfnamefont {S.}~\bibnamefont
  {Muhuri}},\ }\href@noop {} {\bibfield  {journal} {\bibinfo  {journal} {Sci.
  Rep.}\ }\textbf {\bibinfo {volume} {13}},\ \bibinfo {pages} {21391} (\bibinfo
  {year} {2023})}\BibitemShut {NoStop}%
\bibitem [{\citenamefont {Schutz}(2003)}]{schutz-rev}%
  \BibitemOpen
  \bibfield  {author} {\bibinfo {author} {\bibfnamefont {G.~M.}\ \bibnamefont
  {Schutz}},\ }\href@noop {} {\bibfield  {journal} {\bibinfo  {journal} {J.
  Phys. A: Math. Gen.}\ }\textbf {\bibinfo {volume} {36}},\ \bibinfo {pages}
  {R339} (\bibinfo {year} {2003})}\BibitemShut {NoStop}%
\bibitem [{\citenamefont {Muhuri}(2014)}]{driven3}%
  \BibitemOpen
  \bibfield  {author} {\bibinfo {author} {\bibfnamefont {S.}~\bibnamefont
  {Muhuri}},\ }\href@noop {} {\bibfield  {journal} {\bibinfo  {journal}
  {Europhys. Lett.}\ }\textbf {\bibinfo {volume} {106}},\ \bibinfo {pages}
  {28001} (\bibinfo {year} {2014})}\BibitemShut {NoStop}%
\bibitem [{\citenamefont {Chou}\ and\ \citenamefont {Lohse}(1999)}]{chou}%
  \BibitemOpen
  \bibfield  {author} {\bibinfo {author} {\bibfnamefont {T.}~\bibnamefont
  {Chou}}\ and\ \bibinfo {author} {\bibfnamefont {D.}~\bibnamefont {Lohse}},\
  }\href@noop {} {\bibfield  {journal} {\bibinfo  {journal} {Phys. Rev. Lett.}\
  }\textbf {\bibinfo {volume} {82}},\ \bibinfo {pages} {3553} (\bibinfo {year}
  {1999})}\BibitemShut {NoStop}%
\bibitem [{\citenamefont {Muhuri}\ and\ \citenamefont
  {Pagonabarraga}(2010)}]{sm-pre1}%
  \BibitemOpen
  \bibfield  {author} {\bibinfo {author} {\bibfnamefont {S.}~\bibnamefont
  {Muhuri}}\ and\ \bibinfo {author} {\bibfnamefont {I.}~\bibnamefont
  {Pagonabarraga}},\ }\href@noop {} {\bibfield  {journal} {\bibinfo  {journal}
  {Phys. Rev. E}\ }\textbf {\bibinfo {volume} {82}},\ \bibinfo {pages} {021925}
  (\bibinfo {year} {2010})}\BibitemShut {NoStop}%
\bibitem [{\citenamefont {Muhuri}\ \emph {et~al.}(2011)\citenamefont {Muhuri},
  \citenamefont {Shagolsem},\ and\ \citenamefont {Rao}}]{sm-pre2}%
  \BibitemOpen
  \bibfield  {author} {\bibinfo {author} {\bibfnamefont {S.}~\bibnamefont
  {Muhuri}}, \bibinfo {author} {\bibfnamefont {L.}~\bibnamefont {Shagolsem}}, \
  and\ \bibinfo {author} {\bibfnamefont {M.}~\bibnamefont {Rao}},\ }\href@noop
  {} {\bibfield  {journal} {\bibinfo  {journal} {Phys. Rev. E}\ }\textbf
  {\bibinfo {volume} {84}},\ \bibinfo {pages} {031921} (\bibinfo {year}
  {2011})}\BibitemShut {NoStop}%
\bibitem [{\citenamefont {Sugden}\ \emph {et~al.}(2007)\citenamefont {Sugden},
  \citenamefont {Evans}, \citenamefont {Poon},\ and\ \citenamefont
  {Read}}]{fungi1}%
  \BibitemOpen
  \bibfield  {author} {\bibinfo {author} {\bibfnamefont {K.~E.~P.}\
  \bibnamefont {Sugden}}, \bibinfo {author} {\bibfnamefont {M.~R.}\
  \bibnamefont {Evans}}, \bibinfo {author} {\bibfnamefont {W.~C.~K.}\
  \bibnamefont {Poon}}, \ and\ \bibinfo {author} {\bibfnamefont {N.~D.}\
  \bibnamefont {Read}},\ }\href@noop {} {\bibfield  {journal} {\bibinfo
  {journal} {Phys. Rev. E}\ }\textbf {\bibinfo {volume} {75}},\ \bibinfo
  {pages} {031909} (\bibinfo {year} {2007})}\BibitemShut {NoStop}%
\bibitem [{\citenamefont {Muhuri}(2013)}]{fungi2}%
  \BibitemOpen
  \bibfield  {author} {\bibinfo {author} {\bibfnamefont {S.}~\bibnamefont
  {Muhuri}},\ }\href@noop {} {\bibfield  {journal} {\bibinfo  {journal} {EPL}\
  }\textbf {\bibinfo {volume} {101}},\ \bibinfo {pages} {38001} (\bibinfo
  {year} {2013})}\BibitemShut {NoStop}%
\bibitem [{\citenamefont {Shinde}\ \emph {et~al.}(2020)\citenamefont {Shinde},
  \citenamefont {Khan},\ and\ \citenamefont {Muhuri}}]{fungi3}%
  \BibitemOpen
  \bibfield  {author} {\bibinfo {author} {\bibfnamefont {B.}~\bibnamefont
  {Shinde}}, \bibinfo {author} {\bibfnamefont {S.}~\bibnamefont {Khan}}, \ and\
  \bibinfo {author} {\bibfnamefont {S.}~\bibnamefont {Muhuri}},\ }\href@noop {}
  {\bibfield  {journal} {\bibinfo  {journal} {Phys. Rev. Res.}\ }\textbf
  {\bibinfo {volume} {2}},\ \bibinfo {pages} {023111} (\bibinfo {year}
  {2020})}\BibitemShut {NoStop}%
\bibitem [{\citenamefont {Katz}\ \emph {et~al.}(1984)\citenamefont {Katz},
  \citenamefont {Lebowitz},\ and\ \citenamefont {Spohn}}]{KLS-main}%
  \BibitemOpen
  \bibfield  {author} {\bibinfo {author} {\bibfnamefont {S.}~\bibnamefont
  {Katz}}, \bibinfo {author} {\bibfnamefont {J.~L.}\ \bibnamefont {Lebowitz}},
  \ and\ \bibinfo {author} {\bibfnamefont {H.}~\bibnamefont {Spohn}},\
  }\href@noop {} {\bibfield  {journal} {\bibinfo  {journal} {Journal of
  Statistical Physics}\ }\textbf {\bibinfo {volume} {34, Nos. 3/4}},\ \bibinfo
  {pages} {497} (\bibinfo {year} {1984})}\BibitemShut {NoStop}%
\bibitem [{\citenamefont {Pelizzola}\ \emph {et~al.}(2019)\citenamefont
  {Pelizzola}, \citenamefont {Pretti}, ,\ and\ \citenamefont
  {Puccioni}}]{KLS-1}%
  \BibitemOpen
  \bibfield  {author} {\bibinfo {author} {\bibfnamefont {A.}~\bibnamefont
  {Pelizzola}}, \bibinfo {author} {\bibfnamefont {M.}~\bibnamefont {Pretti}}, ,
  \ and\ \bibinfo {author} {\bibfnamefont {F.}~\bibnamefont {Puccioni}},\
  }\href@noop {} {\bibfield  {journal} {\bibinfo  {journal} {Entropy}\ }\textbf
  {\bibinfo {volume} {21(1028)}} (\bibinfo {year} {2019})}\BibitemShut
  {NoStop}%
\bibitem [{\citenamefont {Antal}\ and\ \citenamefont {Schutz}(2020)}]{KLS-2}%
  \BibitemOpen
  \bibfield  {author} {\bibinfo {author} {\bibfnamefont {T.}~\bibnamefont
  {Antal}}\ and\ \bibinfo {author} {\bibfnamefont {G.~M.}\ \bibnamefont
  {Schutz}},\ }\href@noop {} {\bibfield  {journal} {\bibinfo  {journal}
  {arXiv:cond-mat/0001295v1}\ } (\bibinfo {year} {2020})}\BibitemShut {NoStop}%
\bibitem [{\citenamefont {Dhakal}\ and\ \citenamefont
  {Selinger}(2010)}]{nematic}%
  \BibitemOpen
  \bibfield  {author} {\bibinfo {author} {\bibfnamefont {S.}~\bibnamefont
  {Dhakal}}\ and\ \bibinfo {author} {\bibfnamefont {J.~V.}\ \bibnamefont
  {Selinger}},\ }\href@noop {} {\bibfield  {journal} {\bibinfo  {journal}
  {Phys. Rev. E}\ }\textbf {\bibinfo {volume} {81}},\ \bibinfo {pages} {031704}
  (\bibinfo {year} {2010})}\BibitemShut {NoStop}%
\bibitem [{\citenamefont {Solon}\ and\ \citenamefont
  {Tailleur}(2013)}]{active-prl}%
  \BibitemOpen
  \bibfield  {author} {\bibinfo {author} {\bibfnamefont {A.~P.}\ \bibnamefont
  {Solon}}\ and\ \bibinfo {author} {\bibfnamefont {J.}~\bibnamefont
  {Tailleur}},\ }\href@noop {} {\bibfield  {journal} {\bibinfo  {journal}
  {Phys. Rev. Lett.}\ }\textbf {\bibinfo {volume} {111}},\ \bibinfo {pages}
  {078101} (\bibinfo {year} {2013})}\BibitemShut {NoStop}%
\bibitem [{\citenamefont {Dolai}\ and\ \citenamefont {et~al.}(2020)}]{chandan}%
  \BibitemOpen
  \bibfield  {author} {\bibinfo {author} {\bibfnamefont {P.}~\bibnamefont
  {Dolai}}\ and\ \bibinfo {author} {\bibnamefont {et~al.}},\ }\href@noop {}
  {\bibfield  {journal} {\bibinfo  {journal} {Soft Matter}\ }\textbf {\bibinfo
  {volume} {16}},\ \bibinfo {pages} {7077} (\bibinfo {year}
  {2020})}\BibitemShut {NoStop}%
\bibitem [{\citenamefont {Chacko}\ \emph {et~al.}(2024)\citenamefont {Chacko},
  \citenamefont {Muhuri},\ and\ \citenamefont {Tripathy}}]{goutam}%
  \BibitemOpen
  \bibfield  {author} {\bibinfo {author} {\bibfnamefont {J.}~\bibnamefont
  {Chacko}}, \bibinfo {author} {\bibfnamefont {S.}~\bibnamefont {Muhuri}}, \
  and\ \bibinfo {author} {\bibfnamefont {G.}~\bibnamefont {Tripathy}},\
  }\href@noop {} {\bibfield  {journal} {\bibinfo  {journal} {Indian. J. Phys.}\
  }\textbf {\bibinfo {volume} {98}},\ \bibinfo {pages} {1553–1560} (\bibinfo
  {year} {2024})}\BibitemShut {NoStop}%
\end{thebibliography}%
\end{document}